\documentclass[structabstract]{aa}  
\usepackage{amssymb}
\usepackage[british]{babel}
\usepackage{graphicx}
\usepackage{natbib}
\usepackage{rotating}
\usepackage{epsfig}
\usepackage{subfig}
\usepackage{colortbl}
\usepackage[utf8x]{inputenc}
\usepackage{lscape}
\usepackage{tabularx}
\usepackage{longtable}

\usepackage{txfonts}
\begin{document}
\title{Radio spectra and polarisation properties of a bright sample of 
Radio-Loud Broad Absorption Line Quasars.}


\author{G. Bruni \inst{1,2,3}
\and K.-H. Mack \inst{1}
\and E. Salerno \inst{1}  
\and F.M. Montenegro-Montes \inst{4}
\and R. Carballo \inst{5}
\and C.R. Benn \inst{6}
\and J.I. Gonz\'alez-Serrano \inst{7}
\and J. Holt \inst{8}
\and F. Jim\'enez-Luj\'an \inst{3,6,7}}

        
	   
   \institute {INAF-Istituto di Radioastronomia, via Piero Gobetti, 101, I-40127 Bologna, Italy
   \and Università di Bologna, Dip. di Astronomia, via Ranzani, 1, I-40127 Bologna, Italy 
   \and Dpto. de F\'isica Moderna, Univ. de Cantabria, Avda de los Castros s/n, E-39005 Santander, Spain
   \and European Southern Observatory, Alonso de C\'ordova 3107, Vitacura, Casilla 19001, Santiago, Chile
   \and Dpto. de Matem\'atica Aplicada y Ciencias de la Computaci\'on, Univ. de Cantabria, 
ETS Ingenieros de Caminos, Canales y Puertos, Avda de los Castros s/n, E-39005 Santander, Spain
   \and Isaac Newton Group, Apartado 321, E-38700 Santa Cruz de La Palma, Spain
   \and Instituto de F\'isica de Cantabria (CSIC-Universidad de Cantabria), Avda. de los Castros s/n, 
E-39005 Santander, Spain
   \and Leiden Observatory, Leiden University, P.O. Box 9513, NL-2300 RA Leiden, The Netherlands}
                    
             
   \date{}


  \abstract
{
The origin of broad-absorption-line quasars (BAL QSOs) is still an
open issue.  Accounting for $\sim 20\%$ of the QSO population, these
objects present broad absorption lines in their optical spectra
generated from outflows with velocities up to $0.2$ $c$. In this work
we present the results of a multi-frequency study of a well-defined
radio-loud BAL QSO sample, and a comparison sample of radio-loud
non-BAL QSOs, both selected from the Sloan Digital Sky Survey (SDSS).
}
{
We aim to test which of the currently-popular models for the 
BAL phenomenon - `orientation' or 'evolutionary' - best accounts
for the radio properties of BAL quasars.
We also
consider a third model in which BALs are due to polar
jets driven by radiation
pressure.}
{
Observations from 1.4 to 43 GHz have been obtained with the VLA and
Effelsberg telescopes, and data from 74 to 408 MHz have been
compiled from the literature. The spectral indices give
clues to the orientation of these objects, while the determination
of the peak frequency can constrain their age, and test the
evolutionary scenario, in which BAL QSOs are young
QSOs. The fractional polarisation and the rotation measure in part reflect the
local magnetic field strength and particle density.}
   {The fractions of resolved sources 
in the BAL and non-BAL QSO samples are similar (16\% \emph{vs} 12\%). The resolved 
sources in the two samples have similar linear sizes (20 to 400 kpc) and 
morphology.
There is weak evidence that the fraction of variable sources amongst BAL QSOs is lower.
The fractions of candidate
GHz-peaked sources are similar in the two samples
(36$\pm$12\% \emph{vs} 23$\pm$8\%), suggesting that BAL QSOs are not generally 
younger than non-BAL QSOs.
BAL and non-BAL QSOs show a large range of spectral indices, including flat-spectrum and 
steep-spectrum sources, consistent with a broad range of orientations.
There is weak evidence (91\% confidence) that the spectral indices of 
the BAL QSOs are steeper than those of non-BAL QSOs, 
mildly favouring edge-on orientations.
At a higher level of significance ($\ge$97\%), the spectra of BAL QSOs are not 
flatter than those of non-BAL QSOs, which suggests that a polar orientation
is not preferred.
The distributions of fractional polarisation in the two samples are
similar, median values 1-3\%.
The distributions of rotation measure are also similar,
the only outlier being the BAL QSO 1624+37, with an extreme rest-frame
Rotation Measure 
(from the literature) of $-$18350$\pm$570 rad m$^{-2}$.
}
   {}

     \keywords{Quasars: absorption lines - Galaxies: active - Galaxies: evolution - Radio continuum: galaxies}
   \maketitle
%


\section{Introduction}

About 20\% of quasars exhibit broad absorption lines (BALs) in the blue wings of the UV
resonance lines, due to ionised gas with outflow velocities up to
0.2~c (Hewett \& Foltz 2003).   
BALs often obscure parts of the broad emission lines, so the
BAL region must lie outside the broad emission-line region, i.e. $>$
0.1 pc from the quasar nucleus (and probably 10s - 100s pc away). 
For a long time BAL quasars (BAL QSOs) were believed to be
rare amongst luminous radio quasars (Stocke
et al. 1992).  But with the advent of large comprehensive radio surveys it
has become clear that BAL QSOs constitute a significant fraction of the
QSO population (Becker et al. 2000). Becker et al. (2001)
estimated that BAL QSOs are four times less common among quasars
with  $\rm{log} R^{*} > 2$ than among quasars with $\rm{log} R^{*} < 1$, 
where $R^{*}$ is the radio-loudness parameter being defined by Stocke et
al. (1992). 
Hewett \& Foltz (2003) noted that optically-bright BAL QSOs are half
as likely as non-BAL QSOs to have $S_{1.4~\rm{GHz}} > 1$ mJy. 
This rarity has in the past made it
difficult to compile a large sample of radio-loud  
\footnote{The populations of radio-loud and radio-quiet QSOs correspond to radio luminosities of 
$L_{\rm 5~GHz} \ge 10^{26}$ W Hz$^{-1}$ and 
$L_{\rm 5~GHz} < 10^{25}$ W Hz$^{-1}$ respectively (Miller et al. 1990).}
 sample of BAL QSOs which is also radio bright.

There is still no consensus about the origin 
of the absorbing gas in BAL QSOs, the mechanism
which accelerates it, or the
relationship between BAL QSOs and the quasar population as a whole.

Three models have been proposed to explain the presence of BALs: 

(1) In the \emph{orientation model} proposed by \cite{Elvis}, BALs are
produced by a thin-walled funnel-shaped outflow, rising vertically
from a narrow range of radii on the accretion disk and then bending
outward to a cone angle of $\sim60^\circ$ under 
radiation pressure. 
When viewed at certain angles this structure absorbs light from the
QSO nucleus, giving 
rise to BALs. 
In this model, the covering factor of the outflow
would be the same as the observed fraction of BAL QSOs,
i.e. $\sim$20\%.   This model was proposed for radio-quiet QSOs,
since at the time most of the BALs had been found in radio-quiet
QSOs. Elvis (2000) suggests as a possible scenario for the radio-loud
BAL QSOs that the magnetic fields could recollimate the outflow near the
point at which it would otherwise be accelerated radially to BAL
velocities and instead accelerate the flow towards the poles.

(2) On the basis of radio-variability arguments and work
by Punsly (1999a, 1999b), Zhou et al. (2006) and Ghosh \& Punsly (2007)
propose that some BAL QSOs, including both radio-loud and radio-quiet,
are viewed nearly face-on, with the BAL outflows aligned within
15$^{\circ}$ of the polar direction. Punsly (1999b) notes that the
\emph{bipolar wind model} does not preclude the co-existence of equatorial BAL
winds.

(3) In the \emph{evolutionary scheme}, the broad
absorption troughs are produced during a specific period in the
evolution of the quasar, perhaps as it transforms itself from a fully
enshrouded object with a large infrared luminosity, through a BAL
phase, into a normal quasar (e.g., Briggs et al. 1984;
\citealt{Lipari}). BAL QSOs could thus be newborn quasars in which strong
nuclear starburst activity is expelling the dusty
cocoons of the QSOs. This hypothesis finds support in the radio, with a large
fraction of radio-loud BAL QSOs, $\sim$2/3 showing spectral shapes and
morphologies similar to Giga-Hertz-Peaked (GPS) or Compact Steep
Spectrum (CSS) sources (Montenegro-Montes et al. 2008a, MM08
hereafter), a class of radio sources interpreted as either young radio
sources (\citealt{Fanti}) or radio sources frustrated by interaction
with a dense environment (\citealt{Breugel}).

%
%
%
MM08 studied a sample comprising the 15 radio-brightest BAL QSOs
known in 2005, with flux densities $S_{1.4~\rm{GHz}}>15$ mJy. They
measured radio flux densities using both the
100-m Effelsberg telescope and the VLA, over a broad range of frequencies
from 1.4 to 43 GHz. 
Many of the radio characteristics of
these sources were found to be prototypical of CSS or GPS sources. The
low flux-density limit of this sample did not allow MM08 to obtain
significant polarisation measurements, and for only a few sources was
it possible to make VLBI
follow up observations with useful signal-to-noise
(Montenegro-Montes et al. 2008b). 

To overcome these difficulties we
define here a new sample with a brighter flux density limit, 
$S_{\rm 1.4~GHz} > 30$ mJy. 
This sample was obtained by 
correlating the FIRST Catalogue (Faint Images of the Radio Sky at
Twenty-cm; Becker et al. 1995, \citealt{White}) with the
4th SDSS Quasar Catalogue (Schneider et al. 2007) drawn from the 5th data release
of the Sloan Digital Sky Survey (SDSS-DR5; Adelman-McCarthy et
al. 2007). This sample is therefore more homogeneous than the one 
studied in MM08. 

In this paper we report the results of a statistical comparison
between the radio properties of a subsample of radio-loud QSOs showing 
BAL-like features and a matched sample of radio-loud non-BAL QSOs, in order to test
for consistency with the models discussed above. 
In particular, we measure the shape of the synchrotron spectra,
the turn-over frequency and the polarisation properties, for the 
following reasons:

\begin{itemize}

\item	The distribution of radio spectral indices 
constrains the distribution of orientations for a given population
of radio sources (Orr \& Browne 1982), since flatter spectral indices 
imply lines of sight closer to the radio axis.
If the distribution of radio spectral indices of BAL QSOs were
different from that of non-BAL QSOs, this would support
the orientation hypothesis for the origin of BALs.\\

\item The synchrotron turn-over frequency can be used to estimate 
the age of a source, assuming the source is not frustrated 
(recent studies of GPS and CSS sources
tend to exclude the frustration scenario: \citealt{Gupta}, \citealt{Morganti}). 
The age estimate is based on the observed correlation between 
linear size and turnover frequency of CSS and GPS radio sources 
(\citealt{ODea2,Dallacasa}). 
If BAL QSOs are found to be younger
 than the non-BAL QSOs in the comparison sample, the evolutionary hypothesis
 would be favoured.\\

\item Polarisation properties provide clues about magnetic fields and 
particle densities in the environment of the active galactic nucleus.
In particular, if a higher rotation measure is found for BAL QSOs,
this may imply a denser environment.\\

\end{itemize}

The outline of the paper is as follows:
In Section 2 we describe the criteria used to select the BAL QSO sample and the 
non-BAL QSO comparison sample. The radio observations are described in Section 3.
Section 4 presents the results and, for each measured parameter, 
a comparison of the properties of the BAL and non-BAL samples, 
and a discussion of how this comparison affects our view of the
competing hypotheses for the origin of BALs.

The cosmology adopted for the paper assumes a flat universe and the
following parameters: $H_{0}$=70 km s$^{-1}$ Mpc$^{-1}$,
$\Omega_{\Lambda}$=0.7, $\Omega_{M}$=0.3. 
The sign of the quoted spectral indices $\alpha$ is defined by
$S_{\nu} \propto \nu^{\alpha}$.


\section{BAL QSO sample and comparison sample} 

The BAL QSO sample comprises 25 QSOs from the fourth edition of the
SDSS Quasar Catalogue (Schneider et al. 2007) with a FIRST radio
counterpart having $S_{1.4} >$ 30 mJy.  This limit is twice as bright as the one 
used by MM08 in their pilot sample.  We made a two-step selection  of BAL QSOs:
(1) We applied an automatic algorithm using a constant 
continuum to select the QSOs with possible \ion{C}{iv} absorption in their SDSS
spectra. (2) We refined the identification of the BAL QSOs by
interactively fitting the continuum and measuring the absorption index
(AI), as defined by \cite{Trump}, for all the candidate BAL QSOs from
the previous step.  We included in our BAL QSOs sample 
any QSOs with
AI$>$100 km s$^{-1}$.
A comparison sample of non-BAL QSOs is listed
in Table \ref{listsample2}. 
The selection of the samples is described in more detail below.

The fourth edition of the SDSS Quasar Catalogue contains 77429 QSOs,
of which 6226 have a FIRST radio source lying $<$ 2
arcsec away, assumed to be the radio counterpart.
2158 of these lie in the redshift range 1.7$<${\it{z}}$<$4.7, 
allowing identification of \ion{C}{iv} features on SDSS spectra.
We made a two-step search for BALs in the spectra of the 536 of these 
with flux density limit $S_{1.4}>$ 30 mJy. 

The first step in selection of the BAL QSOs was 
measurement of the intrinsic AI as defined by Hall
et al. (2002):
\begin{equation}
{\rm{AI}}=\int_{0}^{25000}(1-\frac{f(v)}{0.9})\cdot Cdv
\end{equation}
where \emph{f(v)} is the normalized flux density. 
The value of $C$ is unity in contiguous intervals of width 450 km s$^{-1}$ 
or greater, over which the quantity in parentheses is everywhere positive; 
otherwise $C=0$. 
AI was computed in \ion{C}{iv} using an automatic procedure in which 
equation 1 was applied using as continuum for the normalization 
the median intensity 
in the rest-frame spectral window 1440-1470 {\AA}.
Objects with $\rm{AI}>0$ were selected as possible BAL QSO candidates.

Each of the 536 spectra were examined by eye
to identify any obvious wrong classifications 
due to, e.g., a poor estimate of the continuum derived from noise peaks 
or other features within the used spectral window. 
As a result of this combined automated and visual 
selection we found 29 initial BAL QSO candidates.

Thirty of the QSOs with AI=0 were randomly selected to
build the control sample, 
with the requirement that their position
in the sky was convenient for scheduling purposes in the
various observing runs, and the distribution in redshift matched as possible the BAL QSO sample. 

These 59 QSOs form the sample of sources 
for which we obtained the multifrequency radio observations for this work.

\begin{figure}[tbp!]
\centering
\includegraphics[width=85mm]{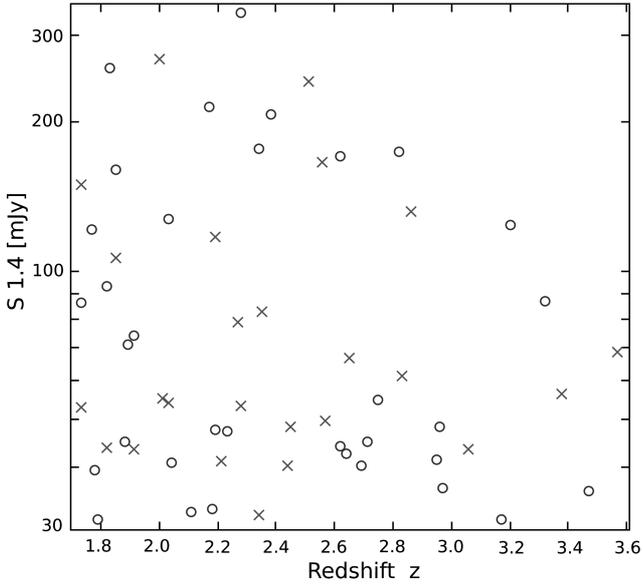}
\caption{Distribution in flux density and redshift, of the QSOs in the BAL (crosses)
and non-BAL (circles) samples.}\label{S}
\end{figure}

\begin{table}
\renewcommand\tabcolsep{3.5pt}
 \centering
\caption{The sample of 25 radio-loud BAL QSOs studied in this paper. 
Columns 2-7 give the optical coordinates and redshifts from SDSS, 
the absorption index for the \ion{C}{iv} line 
(width of at least 1000 km s$^{-1}$), the FIRST peak flux densities and the type.}
\label{listsample1}
  \begin{tabular}{crrrrrc}
  \hline
   \multicolumn{1}{c}{Name}         &
   \multicolumn{1}{c}{RA}           & 
   \multicolumn{1}{c}{DEC}          & 
   \multicolumn{1}{c}{{\it{z}}}            &
   \multicolumn{1}{c}{AI}           &
   \multicolumn{1}{c}{$S_{1.4}$}  &  
   \multicolumn{1}{c}{BAL}           \\
   
   \multicolumn{1}{c}{}             &
   \multicolumn{1}{c}{(J2000)}      & 
   \multicolumn{1}{c}{(J2000)}      &
   \multicolumn{1}{c}{}             &
   \multicolumn{1}{c}{(km s$^{-1}$)}         &
   \multicolumn{1}{c}{(mJy)}    &    
   \multicolumn{1}{c}{Type}             \\     
\hline
\\
0044+00   & 00 44 44.06 & +00 13 03.5  	&   2.28 &    1170$^1$					&   53.1  &	-\\
0756+37   & 07 56 28.24 & +37 14 55.6  	&   2.51 &     330\phantom{$^1$}		&  239.4 &	-\\
0816+48   & 08 16 18.99 & +48 23 28.4  	&   3.57 &     260\phantom{$^1$}		&   68.3  &	-\\
0842+06   & 08 42 24.38 & +06 31 16.8  	&   2.45 &    2690$^1$					&   48.1  	&	-\\
0849+27   & 08 49 14.27 & +27 57 29.7  	&   1.73 &     540\phantom{$^1$}		&   52.8  &	Hi\\
0905+02   & 09 05 52.41 & +02 59 31.5  	&   1.82 &     130\phantom{$^1$}		&   43.5  &	Hi\\
0929+37   & 09 29 13.97 & +37 57 43.0  	&   1.91 &    2170\phantom{$^1$}		&   43.2  &	Hi\\
1014+05   & 10 14 40.35 & +05 37 12.6  	&   2.01 &     250\phantom{$^1$}		&   55.0  &	Hi\\
1040+05   & 10 40 59.80 & +05 55 24.4  	&   2.44 &    4920$^1$					&   40.2  &	-\\
1054+51   & 10 54 16.51 & +51 23 26.1  	&   2.34 &    2220$^1$					&   32.0  &	-\\
1102+11   & 11 02 06.66 & +11 21 04.9  	&   2.35 &     506\phantom{$^1$}		&   82.3  &	-\\
1103+11   & 11 03 34.79 & +11 14 42.4  	&   1.73 &     380\phantom{$^1$}		&  148.0 &	FeLo\\
1129+44   & 11 29 38.47 & +44 03 25.1  	&   2.21 &    1430\phantom{$^1$}		&   41.1  &	Hi\\
1159+01   & 11 59 44.82 & +01 12 06.9  	&   2.00 &    2260$^1$					&  266.5 &	FeLo\\
1159+06   & 11 59 01.75 & +06 56 19.1  	&   2.19 &    3645\phantom{$^1$}		&  116.6 &	Hi\\
1229+09   & 12 29 09.64 & +09 38 10.1  	&   2.65 &     230\phantom{$^1$}		&   66.2  &	-\\
1237+47   & 12 37 17.44 & +47 08 07.0  	&   2.27 &    1300$^1$					&   78.5  &	FeLo\\
1304+13   & 13 04 48.06 & +13 04 16.6  	&   2.57 &     640\phantom{$^1$}		&   49.6  &	-\\
1327+03   & 13 27 03.21 & +03 13 11.2  	&   2.83 &     190\phantom{$^1$}		&   60.7  &	-\\
1335+02   & 13 35 11.90 & +02 53 09.5  	&   1.85 &     210\phantom{$^1$}		&  105.3 & 	Hi\\
1337$-$02 & 13 37 01.40 & $-$02 46 30.3 	&   3.06 &     590\phantom{$^1$}		&   43.3  &	-\\
1404+07   & 14 04 33.01 & +07 28 47.2 		&   2.86 &     120\phantom{$^1$}		&  131.1 &	-\\
1406+34   & 14 06 53.84 & +34 33 37.3  	&   2.56 &     350\phantom{$^1$}		&  164.4 	&	-\\
1603+30   & 16 03 54.15 & +30 02 08.7  	&   2.03 &    1355\phantom{$^1$}		&   53.7  	&	Hi\\
1624+37   & 16 24 53.47 & +37 58 06.6  	&   3.38 &    1020$^1$					&   56.1  	&	-\\

\hline
\end{tabular} 
\begin{list}{}{}
\item[{\bf Notes:}] 
$^1$ Balnicity index $\rm{BI} > 0$ following \cite{Weymann}.
\end{list}

\end{table}

\begin{table}
 \centering
\caption{The sample of 34 comparison (non-BAL) 
QSOs studied in this paper. Columns 2-5 give the 
optical coordinates and redshifts from SDSS 
and the FIRST peak flux densities.}
\label{listsample2}
  \begin{tabular}{crrrr}
  \hline
   \multicolumn{1}{c}{Name}         &
   \multicolumn{1}{c}{RA}           & 
   \multicolumn{1}{c}{DEC}          & 
   \multicolumn{1}{c}{{\it{z}}}            &  
   \multicolumn{1}{c}{$S_{1.4}$}    \\
   
   \multicolumn{1}{c}{}             &
   \multicolumn{1}{c}{(J2000)}      & 
   \multicolumn{1}{c}{(J2000)}      &
   \multicolumn{1}{c}{}             &
   \multicolumn{1}{c}{(mJy)}        \\ 

 
\hline
\\
0014+01    & 00 14 27.93 & +01 13 34.0 		&    2.18 &  32.7       \\
0029$-$09  & 00 29 49.46 & $-$09 51 44.8 	&    2.71 &  44.7       \\
0033$-$00  & 00 33 04.32 & $-$00 48 14.5 	&    1.79 &  31.3     	\\
0103$-$11  & 01 03 28.72 & $-$11 04 14.6 	&    2.19 &  47.5       \\
0124+00    & 01 24 01.75 & +00 35 00.1 		&    1.85 & 159.3      	\\
0125$-$00  & 01 25 17.19 & $-$00 18 29.7 	&    2.28 & 329.1      	\\
0152+01    & 01 52 10.35 & +01 12 28.9 		&    3.17 &  31.3      	\\
0154$-$00  & 01 54 54.37 & $-$00 07 23.1 	&    1.83 & 255.4     	\\
0158$-$00  & 01 58 32.52 & $-$00 42 38.5 	&    2.62 & 169.4      	\\
0750+36    & 07 50 19.55 & +36 30 02.8 		&    2.03 & 126.7      	\\
1005+48    & 10 05 15.98 & +48 05 33.2  	&    2.38 & 206.1  	\\
1322+50    & 13 22 50.55 & +50 03 35.4 		&    1.73 &  85.7     	\\
1333+47    & 13 33 25.06 & +47 29 35.4  	&    2.62 &  44.0  	\\
1401+52    & 14 01 26.15 & +52 08 34.6  	&    2.97 &  36.2  	\\
1411+34    & 14 11 55.24 & +34 15 10.4 		&    1.82 &  92.5      	\\
1411+43    & 14 11 52.77 & +43 00 23.9 		&    3.20 & 122.9      	\\
1502+55    & 15 02 06.53 & +55 21 46.1 		&    3.32 &  86.5       \\
1512+35    & 15 12 58.36 & +35 25 33.3 		&    2.23 &  47.1     	\\
1521+43    & 15 21 49.61 & +43 36 39.4 		&    2.17 & 213.5     	\\			   
1528+53    & 15 28 21.68 & +53 10 30.7 		&    2.82 & 172.4       \\
1554+30    & 15 54 29.40 & +30 01 19.0  	&    2.69 &  40.0  	\\
1634+32    & 16 34 12.77 & +32 03 35.4 		&    2.34 & 175.2     	\\
1636+35    & 16 36 46.41 & +35 57 43.7 		&    1.91 &  73.4     	\\
1641+33    & 16 41 48.07 & +33 45 12.5 		&    2.75 &  54.5      	\\
1728+56    & 17 28 52.61 & +56 41 43.9 		&    1.77 & 120.3      	\\
2109$-$07  & 21 09 26.41 & $-$07 39 25.9 	&    1.88 &  44.8      	\\
2129+00    & 21 29 16.61 & +00 37 56.7  	&   2.96  &  48.0  	\\
2143+00    & 21 43 24.37 & +00 35 02.8	 	&    2.04 &  40.6      	\\
2238+00    & 22 38 43.57 & +00 16 48.0 		&    3.47 &  35.6      	\\
2244+00    & 22 44 59.44 & +00 00 33.4 		&    2.95 &  41.4 	\\
2248$-$09  & 22 48 00.70 & $-$09 07 44.9  	&   2.11  &  32.4  	\\
2331+01    & 23 31 32.84 & +01 06 21.0  	&   2.64  &  42.4  	\\
2346+00    & 23 46 24.56 & +00 19 14.2 		&    1.78 &  39.4      	\\
2353$-$00  & 23 53 30.21 & $-$00 04 13.4 	&    1.89 &  70.5     	\\
\hline
\end{tabular}
\end{table}

\begin{table}
 \centering
  \caption{Summary of the observations.}\label{table_runs}
  \begin{tabular}{cllcc}
  \hline
   Run   &    \multicolumn{1}{c}{Date}  		  &  Telescope  &  Frequencies (GHz)		\\
  \hline
1	&	14$-$23 Dec 07  & Effelsberg 	&2.65, 4.85, 8.35, 10.5		\\
2	&	10$-$15 Sep 08  & Effelsberg 	&2.65, 4.85, 8.35, 10.5		\\
3	&	1$-$5 Jul 09    & Effelsberg 	&2.65, 4.85, 8.35, 10.5		\\
4	&	21$-$27 Jul 09  & VLA(C)     	&1.4, 4.86, 8.46, 22.5, 43.3		\\
\hline
\end{tabular}
\end{table}
\begin{table}
 \centering
 \caption{Observing frequencies and beam sizes (half-power beam-width).}\label{observations}
 \begin{tabular}{cccc}
 \hline
 Telescope & \multicolumn{1}{c}{Frequency} & \multicolumn{1}{c}{Bandwidth} & \multicolumn{1}{c}{$\theta_{\rm{HPBW}}$}  \\ 
           &  \multicolumn{1}{c}{(GHz)}    &  \multicolumn{1}{c}{(MHz)}    & \multicolumn{1}{c}{(arcsec)}     \\ 
\hline  
Effelsberg & 2.65 &     80  & 265    \\
Effelsberg & 4.85 &    500  & 145    \\
Effelsberg & 8.35 &   1100  &  80    \\
Effelsberg & 10.5 &    300  &  65    \\
\hline
VLA(C)     & 1.4  &  50  &  12.5  \\
VLA(C)     & 4.86 &  50  &   3.9  \\
VLA(C)     & 8.46 &  50  &   2.3  \\
VLA(C)     & 22.5 &  50  &   0.9  \\
VLA(C)     & 43.3 &  50  &   0.47 \\
\hline
\end{tabular}
\end{table}

The above selection procedure, normalising the continuum to 
the intensity in a fixed wavelength range, 
is appropriate to deal with large initial samples,  
but has two important caveats: 
(1) AI can be understimated if low-velocity
absorption troughs are superimposed on the emission line, and 
(2) the assumption of a constant 
continuum over the region of interest may be inadequate for some sources. 
As a result, the control sample may include some QSOs which are
more appropriately classified as BAL QSOs, and vice versa.
We measured the AI of the 59 QSOs more accurately using
the following procedure.  
The continuum was obtained by interactively fitting 
the spectral region between the \ion{Si}{iv} to the \ion{C}{iv} emission 
lines (both included) with splines. 
After normalization with the fitted  
continuum, the \ion{C}{iv} AI was measured using Eq. 1 but with the more strict definition 
from Trump et al. (2006), in which parameter $C$ is unity over a 
contiguous interval of 1000 km s$^{-1}$ rather than the 450 km s$^{-1}$ 
used by Hall et al. (2002). 

Of the total 59 QSOs, 25 have AI $>$ 100 km s$^{-1}$ according to this 
definition, and these form the final BAL QSO sample. 
This sample is shown in Table 1, where optical coordinates, redshifts (both from
SDSS), AI, and peak flux densities at 1.4 GHz from FIRST are given.
%
For 11 BAL QSOs with $z \le 2.3$ the SDSS spectra include the vicinity of the 
\ion{Mg}{ii} line at 2798 \AA, allowing a search for 
absorption in \ion{Mg}{ii}, characteristic of low-ionisation BAL
quasars (LoBALs) as well as in \ion{Fe}{ii}  
(rest-frame range from 2200 to 2700 \AA ), characteristic of FeLoBAL quasars.
Three of the sources show \ion{Mg}{ii} absorption as well as \ion{Fe}{ii} absorption 
and are therefore FeLoBALs, labelled `FeLo' in Table 1. 
The remaining 8 sources lack absorption from both \ion{Mg}{ii} and 
\ion{Fe}{ii} and are identified as HiBAL and denoted as `Hi' in Table 1. 
None of the 11 sources are LoBALs,
showing  \ion{Mg}{ii} absorption but no
\ion{Fe}{ii} absorption.

Of the 29 QSOs initially selected as BAL QSOs, 1005+48, 1333+47, 1401+52, 1554+30,
2129+00, 2248$-$09, and 2331+01 have \rm{AI}$<$100 km s$^{-1}$, and were therefore 
included in the comparison sample of non-BAL QSOs. In the opposite sense,   
three QSOs initially in the control sample have \rm{AI}$>$100 km s$^{-1}$ 
and were re-classified as BAL QSOs: 1103+11, 1335+02, and 1404+07. 
The comparison sample is listed in Table 2. 
Figure 1 shows the distribution of the 59 QSOs in redshift and 
$S_{1.4}$ flux density.

We note that the minimum velocity width we used for the 
selection of BAL QSOs, of 1000 km s$^{-1}$, although well above the 
maximum expected values for galactic halos, of $\sim$ 600 km s$^{-1}$, 
is half the value used in the classical definition of BAL QSOs by 
Weymann et al. (1991), which picks up the most extreme cases, and was 
based on radio-quiet QSOs. 
Although a sample of extreme BAL QSOs might better reveal the differences 
between the radio properties of BAL QSOs and a control sample of non-BAL QSOs,
the fraction of BAL QSOs is lower among radio-selected samples than 
among optically selected ones, and moderate BAL QSOs have been included 
in order to have a 
large enough sample for statistical studies. 
In addition, a more relaxed definition of broad absorption allows us to  
cover a wider range of outflow phenomena, including BALs with lower 
outflow velocities. These lower-velocity flows
could be driven by different acceleration mechanisms, depending on the QSO radio luminosity 
(see Punsly 1999a and references therein, and Ghosh \& Punsly 2007 
for BAL QSO models and its relation to radio emission). 
Using Weymann's Balnicity index (\rm{BI}), defined as Eq. 1 apart 
from the lower velocity limit, chosen to be 
3000 km s$^{-1}$, and the above-mentioned wider absorption, 
$ > 2000$ km s$^{1}$, we obtained ${\rm{BI}}>0$ for 7 of the 25 BAL QSOs in this work, 
namely 0044+00, 0842+06, 1040+05, 1054+51, 1159+01, 1237+47 and 1624+37.


\begin{table*}
\centering
\caption{The measured flux densities (in mJy) for the 
sample of 25 radio-loud BAL QSOs.
Flux densities at 1.4, 4.86, 8.46, 22 and 43 GHz are from the VLA; 
those at 2.6, 4.85, 8.35 and 10.5 GHz are from the Effelsberg telescope. 
Asterisked values are taken from the FIRST survey. 
Superscripts on the errors 
indicate the run number of the Effelsberg observations, 
according to the key in Table 3. 
Column 2 the specifies the component of the source being
referred to (see Section 4.1): if no letter is present the total flux 
density is given.
The last column gives
the projected linear size (in kpc) 
of the resolved sources (in boldface) 
and the upper limits for unresolved
sources (the latter taken from the highest-resolution 
map with at least a 3-$\sigma$ detection).}
%
%
\label{listfluxes1}
\renewcommand\tabcolsep{1pt}
\begin{tabular}{lccccccccccr}
\hline
\multicolumn{1}{c}{Name}       &
\multicolumn{1}{c}{comp.}      &
\multicolumn{1}{c}{$S_{1.4}$}  & 
\multicolumn{1}{c}{$S_{2.6}$}  & 
\multicolumn{1}{c}{$S_{4.85}$} &
\multicolumn{1}{c}{$S_{4.86}$} &
\multicolumn{1}{c}{$S_{8.35}$} & 
\multicolumn{1}{c}{$S_{8.46}$} & 
\multicolumn{1}{c}{$S_{10.5}$} & 
\multicolumn{1}{c}{$S_{22}$}   &  
\multicolumn{1}{c}{$S_{43}$}   &
\multicolumn{1}{c}{LS}   \\
\hline
0044+00	 & &	54.76$\pm$0.11*		&54.8$\pm$2.9$^{~1}$ 	&30.9$\pm$1.0$^{~1}$	&	34.2$\pm$1.8	&	24.0$\pm$1.0$^{~1}$	&	25.5$\pm$0.6	&	26.9$\pm$3.9$^{~1}$	&	9.5$\pm$2.7	&	4.9$\pm$1.8	&  $<$8		\\
0756+37	 & &	213$\pm$14\phantom{~~} 	&266.0$\pm$3.1$^{~1}$~~ 	&209.6$\pm$2.6$^{~1}$~~	&	226.2$\pm$2.0~~	&	131.5$\pm$3.0$^{~1}$~~	&	142.1$\pm$1.8~~	&	113.6$\pm$5.2$^{~1}$~~	&	56.2$\pm$2.4~~	&	$<$8.4	&  $<$8		\\
0816+48	 & &	304$\pm$21\phantom{~~} 	&45.8$\pm$1.4$^{~1}$ 	&31.8$\pm$1.0$^{~1}$	&	31.4$\pm$0.5	&	18.4$\pm$0.8$^{~1}$	&	19.5$\pm$0.5	&	15.3$\pm$2.6$^{~1}$	&	$<$5.7	&	-		&  \bf{217}		\\
0842+06	 & &	45.9$\pm$7.5\phantom{~~}	&45.7$\pm$1.9$^{~1}$ 	&30.7$\pm$1.7$^{~1}$	&	29.5$\pm$0.4	&	~~~$<$54$^{~1}$	&	20.9$\pm$0.5	&	15.4$\pm$4.6$^{~1}$	&	6.9$\pm$1.4	&	-	  	&  $<$8		\\
0849+27	 & &	67.08$\pm$0.30*			&49.4$\pm$1.7$^{~1}$ 	&32.0$\pm$1.5$^{~1}$	&	-		&	21.3$\pm$1.0$^{~1}$	&	-		&	17.2$\pm$2.2$^{~1}$	&	-		&	-	 	&  \bf{173-382} 		\\
	 &  A&	~~7.84$\pm$0.15*	 		&-		     	&-			&	-		&	-			&	-		&	-			&	-		&	-	 	&		\\
	 &  B&	~~4.47$\pm$0.15*	 		&-		     	&-			&	-		&	-			&	-		&	-			&	-		&	-	 	&		\\
	 &  C&	53.95$\pm$0.15*	 		&-		     	&-			&	-		&	-			&	-		&	-			&	-		&	-	 	&		\\
	 &  D&	~~0.82$\pm$0.15*	 		&-		     	&-			&	-		&	-			&	-		&	-			&	-		&	-	 	&		\\
0905+02	 & &	32.8$\pm$8.5\phantom{~~}	&87.0$\pm$3.5$^{~2}$	&47.8$\pm$2.7$^{~2}$	&	-		&	27.9$\pm$1.1$^{~2}$	&	28.5$\pm$1.4	&	$<$135$^{~2}$	&	8.8$\pm$1.9	&	-  		&  $<$8		\\
0929+37	 & &	38.9$\pm$4.9\phantom{~~}	&29.4$\pm$1.8$^{~2}$	&29.1$\pm$1.4$^{~2}$	&	32.0$\pm$0.5	&	26.0$\pm$1.3$^{~2}$	&	29.3$\pm$0.5	&	20.7$\pm$7.2$^{~2}$	&	19.1$\pm$1.4~~	&	6.9$\pm$2.5	&  $<$8		\\
1014+05	 & &	57.9$\pm$3.3\phantom{~~}	&37.4$\pm$3.3$^{~3}$	&33.1$\pm$2.2$^{~3}$	&	34.5$\pm$0.6	&	-			&	25.1$\pm$0.5	&	17.9$\pm$3.0$^{~3}$	&	12.6$\pm$2.2~~	&	-	  	&  $<$8		\\
1040+05	 & &	42.20$\pm$0.23*			&-			&-			&	37.2$\pm$1.4	&	-			&	~~5.2$\pm$0.4	&	-			&	-		&	-	  	&  $<$19	\\
1054+51	 & &	38.8$\pm$3.4\phantom{~~}	&32.1$\pm$1.9$^{~1}$	&16.7$\pm$1.1$^{~1}$	&	15.9$\pm$0.5	&	~~8.6$\pm$0.7$^{~1}$	&	~~9.5$\pm$0.6	&	~~7.9$\pm$3.2$^{~1}$	&	$<$12.6	&	-	  	&  $<$19		\\
1102+11	 & &	98.4$\pm$7.9\phantom{~~}	&79.2$\pm$2.4$^{~3}$	&36.4$\pm$2.4$^{~3}$	&	39.8$\pm$0.8	&	18.6$\pm$0.7$^{~2}$	&	19.9$\pm$0.7	&	16.1$\pm$5.0$^{~2}$	&	$<$4.2	&	-	  	&  $<$19		\\
1103+11	 & &	250.43$\pm$0.21*~~		&150.9$\pm$2.2$^{~1}$~~	&95.5$\pm$2.1$^{~1}$	&	100.9$\pm$0.8~~	&	65.9$\pm$1.1$^{~1}$	&	75.0$\pm$0.8	&	58.4$\pm$3.8$^{~1}$	&	43.6$\pm$1.8~~	&	-	  	&  \bf{69}		\\
	 &  A&	76.14$\pm$0.15*			&-			&	-		&	21.2$\pm$0.3	&	-			&	10.2$\pm$0.3	&	-			&	-		&	-	  	&		\\
	 &  C&	174.29$\pm$0.15*~~		&-			&-			&	79.7$\pm$0.7	&	-			&	64.8$\pm$0.7	&	-			&	-		&	-		&		\\
1129+44	 & &	42.1$\pm$7.4\phantom{~~}	&72.1$\pm$1.6$^{~1}$	&41.3$\pm$1.5$^{~1}$	&	38.5$\pm$2.5	&	23.9$\pm$6.6$^{~1}$	&	26.6$\pm$1.7	&	19.5$\pm$3.9$^{~1}$	&	3.1$\pm$1.2	&	-	 	&  $<$19		\\
1159+01	 & &	264.4$\pm$9.8\phantom{~~~~}	&-			&-			&	146.1$\pm$1.3~~	&	-			&	176.6$\pm$2.2~~	&	-			&	169.2$\pm$2.8~~~~	&	-   		&  $<$8		\\
1159+06	 & &	159.5$\pm$3.1\phantom{~~~~}	&109.5$\pm$5.3$^{~2}$~~	&60.2$\pm$2.5$^{~2}$	&	58.5$\pm$1.1	&	35.8$\pm$1.6$^{~2}$	&	37.6$\pm$0.7	&	26.2$\pm$3.3$^{~2}$	&	4.6$\pm$2.3	&	-	  	&  $<$19		\\
1229+09	 & &	81.3$\pm$6.6\phantom{~~}	&35.8$\pm$1.5$^{~1}$	&15.4$\pm$2.3$^{~1}$	&	17.3$\pm$0.5	&	15.7$\pm$6.3$^{~1}$	&	-		&	~~7.5$\pm$2.4$^{~1}$	&	$<$2.7	&	-	  	&  $<$32		\\
1237+47	 & &	74.7$\pm$5.9\phantom{~~}	&-			&-			&	62.3$\pm$1.0	&	-			&	61.6$\pm$1.6	&	-			&	-		&	-	  	&  $<$19	\\
1304+13	 & &	61.4$\pm$4.9\phantom{~~}	&37.0$\pm$3.0$^{~3}$	&21.4$\pm$1.6$^{~3}$	&	24.6$\pm$0.5	&	~~~15$\pm$12~$^{~2}$	&	17.1$\pm$0.5	&	14.3$\pm$3.5$^{~2}$	&	5.3$\pm$1.3	&	-	 	&  $<$8		\\
1327+03	 & &	163$\pm$15\phantom{~~} 	&92.2$\pm$2.7$^{~1}$	&67.7$\pm$1.9$^{~1}$	&	79.5$\pm$1.7	&	52.5$\pm$1.2$^{~1}$	&	56.5$\pm$0.7	&	51.8$\pm$3.3$^{~1}$	&	28.6$\pm$1.8~~	&	-	  	&  $<$8		\\
1335+02	 & &	112.3$\pm$5.6\phantom{~~~~}	&96.9$\pm$1.7$^{~1}$	&81.9$\pm$1.3$^{~1}$	&	86.7$\pm$1.3	&	63.3$\pm$1.4$^{~1}$	&	69.4$\pm$1.0	&	64.7$\pm$3.9$^{~1}$	&	88.6$\pm$1.7~~	&	$<$10.5	&  $<$8			\\
1337$-$02& &	47.2$\pm$5.3\phantom{~~}	&67.3$\pm$3.2$^{~3}$	&42.3$\pm$2.7$^{~3}$	&	42.9$\pm$0.8	&	-			&	16.4$\pm$0.5	&	-			&	-		&	-	  	&  $<$19		\\
1404+07	 & &	187.8$\pm$3.6\phantom{~~~~}	&225.9$\pm$2.6$^{~1}$~~	&209.4$\pm$2.8$^{~1}$~~	&	206.4$\pm$2.3~~	&	173.6$\pm$2.4$^{~1}$~~	&	189.6$\pm$2.2~~	&	160.9$\pm$7.5$^{~1}$~~	&	146.9$\pm$2.3~~~~	&	-	 	&  $<$8		\\
1406+34	 & &	165$\pm$24\phantom{~~} 	&246.6$\pm$2.9$^{~1}$~~	&301.5$\pm$3.1$^{~1}$~~	&	312.8$\pm$3.1~~	&	263.1$\pm$3.4$^{~1}$~~	&	276.3$\pm$3.0~~	&	~237$\pm$10~$^{~1}$	&	180.1$\pm$2.9~~	~~&	90.2$\pm$3.4~~	&  $<$4		\\
1603+30	 & &	54.17$\pm$0.14*			&34.4$\pm$9.9$^{~3}$	&38.8$\pm$2.5$^{~3}$	&	34.5$\pm$2.4	&	-			&	26.9$\pm$0.6	&	-			&	9.7$\pm$1.3	&	-		&  \bf{17}		\\
	 &  A&	-				&-			&-			&	-		&	-			&	-		&	-			&	2.4$\pm$0.8	&	-		&		\\
	 &  C&	-				&-			&-			&	-		&	-			&	-		&	-			&	7.3$\pm$1.0	&	-		&		\\
1624+37	 & &	56.44$\pm$0.14*	 		&35.0$\pm$3.0$^{~1}$	&25.8$\pm$1.5$^{~1}$	&	28.1$\pm$3.7	&	22.5$\pm$0.7$^{~1}$	&	18.2$\pm$1.7	&	18.9$\pm$5.6$^{~1}$	&	7.4$\pm$3.1	&	-		&  $<$19		\\
\hline
\end{tabular}
\end{table*}

\begin{table*}
  \caption{Flux densities for the sample of 34 non-BAL QSOs (see the caption of Table
5 for details). 
For source 2238+00 we give the upper limit of linear size from the FIRST, 
since the only significant detection from our observations 
was at 1.4 GHz, with a lower resolution.}
  \label{listfluxes2}
  \centering
\renewcommand\tabcolsep{1pt}
\begin{tabular}{lccccccccccr}
\hline
\multicolumn{1}{c}{Name}       & 
\multicolumn{1}{c}{comp.}      &
\multicolumn{1}{c}{$S_{1.4}$}  & 
\multicolumn{1}{c}{$S_{2.6}$}  & 
\multicolumn{1}{c}{$S_{4.85}$} & 
\multicolumn{1}{c}{$S_{4.86}$} &
\multicolumn{1}{c}{$S_{8.35}$} & 
\multicolumn{1}{c}{$S_{8.46}$} & 
\multicolumn{1}{c}{$S_{10.5}$} &  
\multicolumn{1}{c}{$S_{22}$}   &  
\multicolumn{1}{c}{$S_{43}$}   &
\multicolumn{1}{c}{LS}   \\
\hline
0014+01	 &   	& 40.9$\pm$6.3\phantom{~~} 		& 	19.6$\pm$1.8$^{~3}$ 	& 13.7$\pm$0.9$^{~3}$ 		& 14.1$\pm$0.5~~ 		& 13.9$\pm$0.7$^{~3}$		& 10.9$\pm$0.4		& ~~8.5$\pm$1.8$^{~3}$		&	4.0$\pm$1.2	&	-	 	&  $<$8		\\
0029$-$09&   	& 41.9$\pm$5.2\phantom{~~}		&	33.5$\pm$2.5$^{~3}$	&	43.5$\pm$2.7$^{~3}$	&	38.6$\pm$1.1~~	&	39.6$\pm$1.0$^{~3}$	&	54.7$\pm$0.7	&	44.4$\pm$2.5$^{~3}$	&	67.9$\pm$2.8~~	&	40.0$\pm$2.2~~	&  $<$4		\\
0033$-$00&   	& 65.1$\pm$5.7\phantom{~~}		&	29.1$\pm$2.3$^{~3}$	&	19.8$\pm$1.3$^{~3}$	&	17.9$\pm$0.7~~	&	12.4$\pm$0.6$^{~3}$	&	13.2$\pm$0.4	&	12.9$\pm$2.2$^{~3}$	&	4.1$\pm$2.0	&	-	  	& \bf{51}	\\
	 &A  	&	-				&-				&	-			&-			&-				&	~~2.5$\pm$0.2	& -				& -			& 	-		&	\\
	 &C  	&	-				&-				&	-			&-			&-				&	10.7$\pm$0.3	& -				& -			& 	-		&	\\
0103$-$11&   	& 50.17$\pm$0.15*			&	48.1$\pm$2.5$^{~3}$	&	41.4$\pm$2.5$^{~3}$	&	40.4$\pm$0.5~~	&	28.6$\pm$0.9$^{~3}$	&	32.2$\pm$0.5	&	26.8$\pm$2.2$^{~3}$	&	24.2$\pm$2.1~~	&	10.4$\pm$1.8~~ 	&  $<$4		\\
0124+00	 &   	&178.4$\pm$6.9\phantom{~~~~}		&	83.9$\pm$3.8$^{~2}$	&	59.2$\pm$2.4$^{~2}$	&	62.8$\pm$1.9~~	&	41.4$\pm$1.5$^{~2}$	&	40.2$\pm$0.8	&	36.7$\pm$4.9$^{~2}$	&	11.6$\pm$1.6~~	&	-	  	&  $<$8	\\
0125$-$00&   	&536.97$\pm$0.42*~~			&	365.7$\pm$3.4$^{~1}~~$	&	248.4$\pm$2.8$^{~1}~~$	&	240.3$\pm$4.5~~~~	&	158.6$\pm$3.7$^{~1}$~~	&	173.7$\pm$1.9~~	&	151.9$\pm$7.7$^{~1}$~~	&	81.2$\pm$2.6~~	&	46.0$\pm$2.9~~	& \bf{42}	\\
	 &A  	&	-				&-				&-				&-			&-				&	27.2$\pm$0.8	& 	-			& 	-		& 	-		&	\\
	 &C  	&	-				&-				&-				&-			&-				&	146.5$\pm$1.7~~	& 	-			& 	-		& 	-		&	\\
0152+01	 &   	& 33.30$\pm$0.15*			&	-			&	-			&	9.2$\pm$1.2	&	-			&	10.7$\pm$0.9	&	-			&	3.3$\pm$1.3	&	$<$1.2	&  $<$19		\\
0154$-$00&   	&273.8$\pm$8.2\phantom{~~~~}		&	234.6$\pm$2.8$^{~2}~~$	&	163.5$\pm$5.9$^{~2}~~$	&	162.3$\pm$1.5~~	&	94.6$\pm$3.1~~$^{~2}$	&	-		&	68.1$\pm$4.3$^{~2}$	&	27.0$\pm$1.7~~	&	7.3$\pm$2.0	&  $<$4		\\
0158$-$00&   	&190.9$\pm$4.4\phantom{~~~~}		&	106.4$\pm$3.2$^{~2}~~$	&	75.1$\pm$2.8$^{~2}$	&	75.0$\pm$0.7	&	51$\pm$19$^{~~2}$	&	52.6$\pm$0.7	&	45.6$\pm$3.6$^{~2}$	&	-		&	12.4$\pm$1.9~~ 	&  $<$4		\\
0750+36	 &   	&109.9$\pm$5.1\phantom{~~~~}		&	-			&	66.1$\pm$2.6$^{~2}$	&	65.4$\pm$0.8	&	45.2$\pm$1.2~~$^{~2}$	&	46.9$\pm$0.7	&	37.5$\pm$3.6$^{~2}$	&	20.5$\pm$2.1~~	&	$<$5.1	&  $<$8		\\
1005+48	 &   	&161$\pm$22\phantom{~~}		&	162.2$\pm$2.5$^{~1}~~$	&	111.5$\pm$4.4$^{~1}$~~	&	121.3$\pm$1.0~~	&	69.9$\pm$3.2~~$^{~1}$	&	83.6$\pm$1.1	&	68.7$\pm$4.0$^{~1}$	&	45.7$\pm$1.6~~	&	10.9$\pm$2.4~~	&  $<$4		\\
1322+50	 &   	&116$\pm$10\phantom{~~}		&	78.9$\pm$2.4$^{~1}$	&	54.3$\pm$1.5$^{~1}$	&	55.6$\pm$0.9	&	42.7$\pm$1.1~~$^{~1}$	&	40.2$\pm$0.7	&	35.6$\pm$3.2$^{~1}$	&	14.3$\pm$2.6~~	&	-	  	&  $<$8	\\
1333+47	 &   	&176.4$\pm$6.9\phantom{~~~~}		&	-			&	26.6$\pm$1.7$^{~3}$	&	25.9$\pm$0.5	&	-			&	17.8$\pm$0.5	&	-			&	13.9$\pm$3.5~~	&	-		&  $<$8	\\
1401+52	 &   	&100$\pm$13\phantom{~~}		&	47.8$\pm$3.3$^{~3}$	&	36.7$\pm$2.4$^{~3}$	&	33.2$\pm$0.6	&	-			&	24.9$\pm$1.2	&	-			&	8.3$\pm$1.6	&	-		&  $<$8	\\
1411+34	 &   	&189.14$\pm$0.24*~~			&	145.0$\pm$2.4$^{~1}~~$	&	120.2$\pm$1.9$^{~1}$~~	&	85.1$\pm$5.2	&	105.1$\pm$1.9~~$^{~1}$~~	&	86.3$\pm$2.9	&	102.6$\pm$5.3$^{~1}$~~	&	120.2$\pm$1.7~~~~	&	83.3$\pm$2.7~~  	& \bf{199}	\\
	 &A  	& 28.32$\pm$0.14*			&	-			&	-			&	~~5.4$\pm$0.7	&	-			&	~~2.6$\pm$0.7	&	-			&-			&	-		&	\\
	 &B  	& 58.17$\pm$0.14*			& 	-			&	-			&	16.4$\pm$0.7	&	-			&	~~5.3$\pm$0.7	&-				&	-		&	-	 	&	\\
	 &C  	&102.65$\pm$0.14*~~			&	-			&	-			&	80.4$\pm$1.0	&	-			&	85.5$\pm$1.3	&	-			&	-	 	&	-	 	&	\\
1411+43	 &   	&135.4$\pm$8.9\phantom{~~~~}		&	112.1$\pm$2.0$^{~1}~~$	&	98.6$\pm$1.4$^{~1}$	&	99.9$\pm$0.9	&	79.3$\pm$1.2~~$^{~1}$	&	82.3$\pm$1.1	&	73.5$\pm$4.1$^{~1}$	&	63.7$\pm$2.0~~	&	29.6$\pm$3.0~~ 	&  $<$4	\\
1502+55	 &   	& 91.94$\pm$0.15*			&	58.9$\pm$1.6$^{~1}$	&	35.2$\pm$1.1$^{~1}$	&	35.6$\pm$0.7	&	~~~34$\pm$16~~~$^{~1}$	&	24.9$\pm$0.5	&	19.7$\pm$2.7$^{~1}$	&	7.2$\pm$1.7	&	-	  	&  $<$8	\\
1512+35	 &   	& 47.48$\pm$0.13*			&	40.8$\pm$2.2$^{~2}$	&	36.0$\pm$1.9$^{~2}$	&	40.9$\pm$0.5	&	34.7$\pm$4.7~~$^{~2}$	&	48.4$\pm$0.8	&	32.8$\pm$3.4$^{~2}$	&	60.4$\pm$2.5~~	&	16.8$\pm$2.9~~ 	&  $<$4	\\
1521+43	 &   	&268.0$\pm$4.5\phantom{~~~~}		&	276.0$\pm$2.7$^{~1}~~$	&	256.8$\pm$2.8$^{~1}~~$	&	387.3$\pm$3.0~~	&	346.9$\pm$4.4~~$^{~1}$~~	&	523.1$\pm$5.7~~	&	~~~~396$\pm$16~$^{~1}$~~~	&	583.7$\pm$6.2~~~~	&	-	 	&  $<$8	\\
1528+53	 &   	&207$\pm$12\phantom{~~}		&	131.7$\pm$8.4$^{~1}~~$	&	68.0$\pm$1.4$^{~1}$	&	70.6$\pm$3.7	&	45.0$\pm$1.5~~$^{~1}$	&	43.0$\pm$0.8	&	36.3$\pm$3.5$^{~1}$	&	15.1$\pm$1.5~~	&	-	 	&  $<$8	\\
1554+30	 &   	& 41.22$\pm$0.15*			&	31.4$\pm$2.7$^{~3}$	&	44.6$\pm$2.8$^{~3}$	&	37.1$\pm$1.9	&	-			&	30.3$\pm$2.2	&	-			&	10.1$\pm$2.9~~	&	2.6$\pm$1.2	&  $<$8\\
1634+32	 &   	& 46.9$\pm$1.6\phantom{~~}		&	209.9$\pm$2.8$^{~2}~~$	&	171.6$\pm$5.8$^{~2}~~$	&	179.3$\pm$1.5~~	&	136.5$\pm$2.9~~$^{~2}$~~	&	148.3$\pm$1.9~~	&	116.9$\pm$5.7$^{~2}$~~	&	35.1$\pm$3.3~~	&	13.9$\pm$2.6~~	&  $<$4	\\
1636+35	 &   	& 86.6$\pm$7.0\phantom{~~}		&	97.7$\pm$2.5$^{~2}$	&	76.1$\pm$2.9$^{~2}$	&	-		&	58.1$\pm$2.9~~$^{~2}$	&	-		&	48.8$\pm$3.9$^{~2}$	&	29.3$\pm$2.0~~	&	-	 	&  $<$8	\\
1641+33	 &   	& 83.7$\pm$6.4\phantom{~~}		&	73.1$\pm$2.2$^{~2}$	&	73.4$\pm$2.8$^{~2}$	&	81.4$\pm$0.8	&	70.6$\pm$1.7~~$^{~2}$	&	66.1$\pm$0.9	&	61.7$\pm$3.6$^{~2}$	&	26.6$\pm$3.3~~	&	-	  	&  $<$8	\\
1728+56	 &   	&190$\pm$14\phantom{~~}		&	108.6$\pm$2.7$^{~1}~~$	&	56.4$\pm$1.3$^{~1}$	&	67.2$\pm$0.9	&	34.9$\pm$1.7~~$^{~1}$	&	35.8$\pm$1.4	&	25.3$\pm$3.3$^{~1}$	&	15.4$\pm$1.6~~	&	-	  	& \bf{87}	\\
	 &A  	&	-				&	-			&		-		&	26.6$\pm$0.4	&	-			&	14.9$\pm$0.3	&	-			&	4.2$\pm$1.1	&	-	  	&	\\			
	 &B  	&	-				&	-			&	-			&	-		&	-			&	-		&	-			&	4.9$\pm$0.8	&	-		&	\\
	 &C  	&	-				&	-			&	-			&	-		&	-			&	-		&	-			&	6.3$\pm$0.9	&	-		&	\\
	 &B+C	&	-				&	-			&		-		&	40.9$\pm$0.5	&	-			&	22.3$\pm$0.4	&	-			&	-		&	-	  	&	\\
2109$-$07&   	& 59.5$\pm$3.2\phantom{~~}		&	41.7$\pm$2.7$^{~3}$	&	33.7$\pm$2.2$^{~3}$	&	30.9$\pm$0.7	&	19.9$\pm$0.8~~$^{~3}$	&	22.6$\pm$0.6	&	15.1$\pm$2.8$^{~3}$	&	-		&	-	 	& $<$19	\\
2129+00	 &	& 51.4$\pm$7.6\phantom{~~}		&	33.2$\pm$1.5$^{~1}$	&	30.4$\pm$3.0$^{~1}$	&	32.8$\pm$0.5	&	24.2$\pm$0.8~~$^{~1}$	&	27.4$\pm$0.5	&	21.6$\pm$2.6$^{~1}$	&	10.3$\pm$1.6~~	&	8.6$\pm$2.3	&  $<$8	\\
2143+00	 &	& 41.45$\pm$0.10*			&	38.5$\pm$2.4$^{~3}$	&	53.9$\pm$3.3$^{~3}$	&	50.9$\pm$0.6	&	53.6$\pm$1.3~~$^{~3}$	&	61.9$\pm$0.8	&	56.9$\pm$2.2$^{~3}$	&	39.4$\pm$1.9~~	&	26.3$\pm$1.8~~ 	&  $<$4	\\
2238+00	 &	& 32.2$\pm$7.9\phantom{~~}		&	-			&	-			&	-		&	-			&	-     		&	-	     		&	$<$5.4	&	- 		&  $<$37	\\
2244+00	 &	& 34.4$\pm$7.1\phantom{~~}		&	-			&	-			&	24.9$\pm$0.7	&	-			&	22.2$\pm$0.5	&	-			&	8.9$\pm$1.5  	&	4.9$\pm$1.5	&  $<$8	\\
2248$-$09&	& 35.33$\pm$0.13*			&	19.4$\pm$1.6$^{~1}$	&	14.9$\pm$2.5$^{~1}$	&	13.2$\pm$0.4	&	~~7.9$\pm$0.7~~$^{~1}$	&	~~8.1$\pm$0.5	&	~~7.4$\pm$2.0$^{~1}$	&	-		&	-	  	& $<$19	\\
2331+01	 &	& 41.96$\pm$0.13*			&	38.4$\pm$2.0$^{~1}$	&	28.5$\pm$2.5$^{~1}$	&	25.9$\pm$0.5	&	19.3$\pm$0.6~~$^{~1}$	&	23.0$\pm$0.5	&	17.9$\pm$2.3$^{~1}$	&	10.7$\pm$1.4~~	&	$<$7.5	&  $<$8	\\
2346+00	 &	& 48.4$\pm$2.9\phantom{~~}		&	41.9$\pm$2.5$^{~2}$	&	47.6$\pm$2.1$^{~2}$	&	49.7$\pm$0.6	&	41.8$\pm$4.0~~$^{~2}$	&	50.1$\pm$0.8	&	42.7$\pm$3.8$^{~2}$	&	37.1$\pm$1.9~~	&	16.7$\pm$2.6~~	&  $<$4	\\
2353$-$00&	& 87$\pm$17\phantom{}		&	-			&	$<$63$^{~2}$	&	26.8$\pm$0.6	&	-			&	18.5$\pm$0.5	&	$<$29$^{~2}$	&	7.9$\pm$1.8	&	1.9$\pm$0.7   	&  $<$8	\\
\hline
\end{tabular}
\end{table*}


\section{Radio observations and data reduction}

We observed the QSOs at frequencies ranging from 1.4 to 43 GHz, 
using the 100-m Effelsberg single dish and the VLA in full polarisation mode 
(Stokes I, Q and U images). Tables 3 and 4 summarise the different runs and observing
setups.

\subsection{Effelsberg 100-m telescope}

Observations with the Effelsberg 100-m dish were carried out during 3
separate runs (see Table \ref{table_runs}). 
All observations (for BAL QSOs
and comparison QSOs) were carried out using cross-scans in azimuth and
elevation at 2.65, 4.85, 8.35 and 10.5 GHz, with a cross-scan length
of 4 times the beam size. On-source integration times were between 20
and 60 seconds per source and per frequency, depending on the expected
source intensity.

During the data reduction, all scans were visually
checked to remove radio-frequency interference, bad-weather effects
(noisy scans due to heavy rain or clouds) or detector
instabilities. The signals were fitted with a Gaussian to extract flux
densities, following the standard reduction method for Effelsberg
data, using the CONT2 programme of the
TOOLBOX\footnote{http://www.mpifr-bonn.mpg.de/english/radiotelescope/index.html}
package.
3-$\sigma$ upper limits were placed on the flux densities of undetected
sources (Section 3.3).

The flux-densities were calibrated on the Baars et al. (1977) scale,
via observations of 3C286.  
A calibration of the polarisation was carried out in the standard way,
using observations of 3C286
to remove the effects of instrumental polarisation.

\subsection{Very Large Array}

In July 2009 we used the VLA to observe
the BAL QSOs in the frequency range 1.4 to 43 GHz. We used five
different receivers (L, C, X, K and Q band, corresponding to 1.4,
4.86, 8.46, 22.5 and 43.3 GHz) and imaged
at all frequencies the QSOs in both 
the BAL QSO sample and the comparison sample.  The 
integration times depended on band and source, and varied beween 4
and 15 minutes.

The highest angular resolutions reached in our work are lower than
those reached by MM08, since the latter used VLA in A configuration
whereas that for this work we used configuration C at the VLA.

The flux-densities were calibrated on the Baars et al. (1977) scale,
via observations of 3C286.  
3C286 was also used as a phase
calibrator. 
In addition, secondary phase calibrators were
observed at regular time intervals (different for each band) and 
between 2 and 5 degrees from the target sources. 
At the highest frequencies (22 and 43 GHz) we
switched between target and calibrator
every 30 seconds (`fast-switching mode'), to improve the phase calibration.

The data were reduced with the 31DEC09 version of
AIPS\footnote{http://www.aips.nrao.edu}, and version 3.0 of
CASA\footnote{http://casa.nrao.edu} was used to extract flux densities
via an automated python script. 
We used the task IMSTAT to perform
this calculation for the Stokes I, Q and U images. 
3-$\sigma$ upper limits were placed on the flux densities of undetected
sources (Section 3.3).

The polarisations were calibrated using 3C286 as a strong
unresolved source to determine the instrumental polarisation and the
apparent polarisation angle.

\subsection{Error determination}

We followed the approach of \cite{Klein} for the determination of flux-density 
errors, 
considering three main contributions:
(i) the fractional calibration error $\Delta S_{cal}$, 
estimated from the dispersion of the observations of the flux density 
calibrators; 
(ii) the error introduced by noise, $\Delta S_{n}$, 
which is estimated from the local noise around the 
source; 
and 
(iii) the confusion error $\Delta S_{conf}$ due to 
the possible presence of background sources within the beam area. 
When the beam has small dimensions, as in interferometric data, 
the last term can be neglected.

%
Thus the expressions for the total uncertainty of the Stokes parameters are given by 
equations \ref{err} and \ref{err2} below for 
Effelsberg and the VLA, respectively:
\begin{equation}\label{err}
\Delta S_{i}= \sqrt{(S_{i} \cdot \Delta S_{cal})^{2} + \Delta S_{n,i}^{2} + \Delta S_{conf,i}^{2}} \hspace{0.5cm} i={\rm I,Q,U}
\end{equation}
\begin{equation}\label{err2}
\Delta S_{i}= \sqrt{(S_{i} \cdot \Delta S_{cal})^{2} + \Delta S_{n,i}^{2} \cdot \frac{A_{src}}{A_{beam}}} \hspace{0.5cm} i={\rm I,Q,U}
\end{equation}
where $A_{src}$ is the area of the aperture within which the source flux
density is measured, and $A_{beam}$ is the area of the
synthesised beam. 
The expressions for the
uncertainties of $m$ (fractional polarisation) and $\chi$
(polarisation angle) can be found in \cite{Klein}.
%


\section{Results and discussion}


In Tables \ref{listfluxes1} and \ref{listfluxes2} we present our 
measurements of the flux-densities
of the BAL QSOs and the comparison sample from 1.4 to 43 GHz. 
For sources which were resolved on the VLA maps (see Section 4.1) we provide both 
the total flux densities and the flux densities of the individual components, 
where the components were well resolved and when reliable measurements were possible. 
At 1.4 GHz, if we made no measurements,
we give in Tables \ref{listfluxes1} and \ref{listfluxes2}
the FIRST integrated flux densities.
In the last column of each table we give upper limits on the projected linear sizes
 of the unresolved sources, extracted from the 
highest-resolution VLA map with a significant detection.
The sizes of resolved sources were obtained from whichever
 map presents the largest projected linear size.

We extended the SEDs to frequencies lower than 1.4 GHz using 
data from the literature (see Table \ref{LF1}). 
Flux densities and upper limits 
(when cut-out images were available) were collected from the following surveys: 
VLSS (74 MHz, \citealt{Cohen}), 6C (151 MHz, \citealt{Hales}), 
WENSS (325 MHz, \citealt{Bruyn}), TEXAS (365 MHz, \citealt{Douglas}) 
and B3 (408 MHz, \citealt{Ficarra}).

\begin{table}
 \centering
  \caption{Low-frequency flux densities (in mJy) 
from the literature (see Section 4 for references) 
for the samples of BAL (upper list) and non-BAL (bottom list) 
QSOs.  The subscripts indicate the frequencies in MHz. 
For VLSS and WENSS, 3-$\sigma$ 
upper limits are given for non-detections.}
  \label{LF1}
\renewcommand\tabcolsep{1pt}
  \begin{tabular}{ccccccc}
  \hline
   \multicolumn{1}{c}{}             &
   \multicolumn{1}{c}{}             & 
   \multicolumn{1}{c}{}             &
   \multicolumn{1}{c}{}             &
   \multicolumn{1}{c}{}             & 
   \multicolumn{1}{c}{}             \\
 
   \multicolumn{1}{c}{Name}               &
   \multicolumn{1}{c}{$S^{VLSS}_{74}$}    & 
   \multicolumn{1}{c}{$S^{6C}_{151}$}     &   
   \multicolumn{1}{c}{$S^{WENSS}_{325}$}  & 
   \multicolumn{1}{c}{$S^{TEXAS}_{365}$}  &
   \multicolumn{1}{c}{$S^{B3}_{408}$}     \\
   \multicolumn{1}{c}{}             &
   \multicolumn{1}{c}{}             & 
   \multicolumn{1}{c}{}             &
   \multicolumn{1}{c}{}             &
   \multicolumn{1}{c}{}             & 
   \multicolumn{1}{c}{}             \\ 
\hline
0044+00    	&$<$300~~~~~~~~~  	&-		&-   			&-  		& -	      	\\  
0756+37    	&$<$228~~~~~~~~~ 	&-		&25.0$\pm$3.1   	&- 		&-	      	\\  
0816+48    	&$<$936~~~~~~~~~	&-		&62.7$\pm$4.5   	&-		&-	      	\\  
0842+06    	&$<$267~~~~~~~~~  	&-		&-   	 		&-  		&-	      	\\  
0849+27    	&453$\pm$75		&-		&-   			&-  		&-	      	\\	   
0905+02    	&$<$360~~~~~~~~~ 	&-		&-   			&-  		&-	      	\\	   
0929+37    	&$<$909~~~~~~~~~ 	&-		&91.0$\pm$3.2   	&-  		&100$\pm$20   	\\  
1014+05    	&$<$489~~~~~~~~~ 	&-		&-   			&-  		&-	      	\\	   
1040+05    	&$<$405~~~~~~~~~ 	&-		&-   			&-  		&-	      	\\  
1054+51    	&$<$201~~~~~~~~~  	&-		&42.5$\pm$3.6  		&-  		&-	      	\\	   
1102+11    	&$<$297~~~~~~~~~ 	&-		&-  			&-  		&-	      	\\	   
1103+11    	&4050$\pm$420		&-		&-   			&1091$\pm$61~~  	&810$\pm$50   	\\ 	 
1129+44    	&$<$246~~~~~~~~~  	&-		&$<$13.2~~~~~~  		&-  		&-	      	\\  
1159+01    	&$<$320~~~~~~~~~ 	&-		&418$\pm$5.1		&492$\pm$42 	&-	      	\\
1159+06    	&1140$\pm$140		&-		&- 			&481$\pm$26 	&-	      	\\
1229+09    	&$<$1032~~~~~~~~~~~ 	&-		&-   			&-  		&-	      	\\ 
1237+47    	&$<$243~~~~~~~~~  	&-		&42.5$\pm$2.9~~ 		&-		&-	      	\\	   
1304+13    	&$<$594~~~~~~~~~  	&-		&-   			&-  		&-	      	\\  
1327+03    	&$<$288~~~~~~~~~  	&-		&-   			&-  		&-	      	\\  
1335+02    	&$<$261~~~~~~~~~  	&-		&-   			&- 		&-     	      	\\ 	 
1337$-$02  	&$<$210~~~~~~~~~ 	&-		&-  			&-  		&-	      	\\  
1404+07    	&$<$309~~~~~~~~~ 	&-		&-   			&-  		&-	      	\\ 	 
1406+34    	&$<$318~~~~~~~~~ 	&-		&184.0$\pm$3.6~~~~  	&-  		&-	      	\\  
1603+30    	&$<$225~~~~~~~~~  	&-		&~~33$\pm$4.4 		&-  		&-	      	\\  
1624+37    	&$<$355~~~~~~~~~  	&-		&~~59$\pm$4.0 		&-  		&-	      	\\ 
&&&&&\\
\hline
&&&&&\\
0014+01   	&$<$303~~~~~~~~~ 	&-		&-   			&- 		& -	        \\ 	       
0029$-$09 	&$<$402~~~~~~~~~  	&-		&-   			&-  		& - 	        \\ 	 
0033$-$00 	&562$\pm$95  	&-		&-   			&- 		& -       	\\ 	 
0103$-$11 	&$<$288~~~~~~~~~ 	&-		&-   			&-  		& - 	        \\ 	 
0124+00   	&1140$\pm$140   &-		&-   			&905$\pm$56   	&	-       \\ 	 
0125$-$00 	&4890$\pm$530 	&-		&-   			&1710$\pm$110 	&1590$\pm$90    \\ 	 
0152+01   	&$<$267~~~~~~~~~  	&-		&-   			&- 		& - 	        \\ 	 
0154$-$00 	&$<$420~~~~~~~~~  	&-		&-   			&-  		& -	        \\ 	 
0158$-$00 	&1730$\pm$210  	&-		&-   			&754$\pm$23   	&	-       \\ 	 
0750+36   	&$<$231~~~~~~~~~  	&-		&24.9$\pm$3.4   	&-  		&	-       \\ 	 
1005+48    	&~~760$\pm$120	&630$\pm$50	&-   			&492$\pm$42 	&-	        \\  
1322+50   	&~~960$\pm$120  	&520$\pm$40	&-   			&421$\pm$50   	& -	        \\ 	 
1333+47    	&$<$246~~~~~~~~~  	&-		&34.3$\pm$3.0   	&-  		&-	        \\	
1401+52    	&$<$321~~~~~~~~~ 	&-		&$<$10.2~~~~~~~~~~		&-  		&-	        \\  
1411+34   	&1820$\pm$210 	&830$\pm$90	&453.6$\pm$3.3~~  	&360$\pm$25   	& -	        \\ 	 
1411+43   	&$<$237~~~~~~~~~  	&-		&154.7$\pm$2.8~~  	&-  		&170$\pm$20     \\ 	 
1502+55   	&1480$\pm$170  	&710$\pm$40	&349.6$\pm$3.1~~ 		&340$\pm$19   	&504$\pm$70     \\ 	 
1512+35   	&$<$333~~~~~~~~~		&-		&42.3$\pm$3.5  		&-  		&-	        \\ 	 
1521+43   	&$<$249~~~~~~~~~  	&-		&117.8$\pm$3.6~~  	&-  		&120$\pm$20     \\ 	 
1528+53   	&1250$\pm$150  	&1020$\pm$40~~	&504.8$\pm$4.0~~  	&534$\pm$30   	&	-       \\ 	 
1554+30    	&$<$219~~~~~~~~~ 	&-		&30.1$\pm$4.2   	&-  		&-	        \\	   
1634+32   	&$<$225~~~~~~~~~  	&-		&81.8$\pm$4.0   	&-  		&	-       \\ 	 
1636+35   	&$<$219~~~~~~~~~  	&-		&$<$12.6~~~~~~~~~~    		&-  		&	-       \\ 	 
1641+33   	&$<$195~~~~~~~~~  	&-		&106.1$\pm$4.6~~  	&-  		&	-       \\ 	 
1728+56   	&1720$\pm$220 	&1250$\pm$40~~	&726.9$\pm$3.9~~  	&734$\pm$31   	& -  	        \\ 	 
2109$-$07 	&$<$354~~~~~~~~~ 	&-		&-    			&-  		&	-       \\ 	 
2129+00    	&$<$273~~~~~~~~~  	&-		&-  			&-  		&-	        \\
2143+00   	&$<$390~~~~~~~~~  	&-		&-    			&-  		& -	        \\ 	 
2238+00   	&$<$351~~~~~~~~~  	&-		&-    			&- 		&	-       \\ 	 
2244+00   	&$<$333~~~~~~~~~  	&-		&-    			&-  		&	-       \\ 	 
2248$-$09  	&$<$252~~~~~~~~~  	&-		&-   			&-  		&-	        \\  
2331+01    	&$<$288~~~~~~~~~  	&-		&-   			&-  		&-	        \\     
2346+00   	&$<$384~~~~~~~~~  	&-		&-    			&-  		&	-       \\ 	 
2353$-$00 	&~~900$\pm$130 	&-		&-    			&-  		&	-       \\
\hline
\end{tabular} 
\end{table}


\subsection{Morphology}

The radio morphologies of the QSOs in the two samples can be compared
at the arcsec scale using the VLA maps, which have at 
all frequencies a higher resolution than the Effelsberg 100-m 
single-dish cross-scans. 
At 1.4 GHz, we complement our VLA data with those from 
FIRST data, obtained with a higher resolution. 
Maps of the resolved sources are shown in Fig. \ref{morphology} and their linear 
sizes are listed in Table \ref{listfluxes1} and \ref{listfluxes2}.

\subsubsection*{BAL QSO sample}

Amongst the BAL QSO sample, 16 were detected at high angular resolution 
(3 with resolution 0.47 arcsec, from 43-GHz observations,
and 15 with resolution 0.9 arcsec, from 22-GHz observations).
Only one of them (1603+30) 
was resolved at either of these frequencies. 
Another five sources were observed with a poorer resolution, of
2.3 arcsec (8.46 GHz). 
For 0849+27, lacking VLA observations from our work, the FIRST data provide 
the best resolution.
A detailed discussion of the four resolved BAL QSOs is presented below.


0816+48 is elongated to the south-west in the 1.4-GHz map. 
A gaussian fit yields an angular size of 29 arcsec along the 
major axis and an upper limit of 14 arcsec along the minor axis, 
corresponding to a projected linear size of 217 kpc and $<$105 kpc respectively. 
The total spectral index in the range from 4.86 to 8.46 GHz is 
$\alpha_{4.86}^{8.46} = -0.86\pm0.05$ (see Section 4.4 and Table \ref{listalpha1}).

0849+27, for which we have no usable VLA observations,
is resolved in the FIRST map (resolution of 5 arcsec, see Fig. 2).  
The map shows three components, located at 25 arcsec NE (D), 34 arcsec NE (B) 
and 20 arcsec SW (A) from the core (C), which is coincident with the 
QSO optical position.  
The four components are included in the FIRST catalogue of radio sources. 
Throughout the paper, we will label core components as 'C'.
The largest separation between components is approximately 44 arcsec, 
corresponding to a projected linear size of 382 kpc. The total spectral index 
of the source is $\alpha_{4.86}^{8.46} = -0.73\pm0.12$. 
If the two fainter and farther located NE components are 
interpreted as a background source, 0849+27 would have a size of 20
arcsec, corresponding to a projected linear size of 173 kpc. 

1103+11 is resolved at 1.4 (FIRST), 4.86 and 8.46 GHz, and shows a
core-lobe morphology. 
This interpretation is supported by the coincidence of component 
C with the optical position of the QSO. The angular size from the 
highest-resolution map (8.46 GHz) is 8 arcsec, corresponding to a projected linear size 
of 69 kpc. The lobe is not detected at higher frequencies, probably 
due to the steeper spectral index with respect to the core, which decreases the lobe 
flux density below the 3-$\sigma$ detection limit at the highest frequencies. 
The spectral indices are: $\alpha_{4.86}^{8.46}=-0.37\pm0.02$ for the core (C) 
and $\alpha_{4.86}^{8.46}=-1.32\pm0.06$ for the lobe (A), and $\alpha_{4.86}^{8.46}=-0.53\pm0.02$
for the total emission.

1603+30 is resolved at 22 GHz, showing a core component (C) coincident 
with the optical position of the QSO and another component 
towards the south (A), which could be interpreted as a lobe. 
The angular size is 2 arcsec, corresponding to 17 kpc. 
The total spectral index of the source is $\alpha_{4.86}^{8.46}=-0.45\pm0.13$.

\subsubsection*{non-BAL QSO sample}

For 31 of the 34 sources in the comparison sample, we obtained observations
with resolution 0.5 or 0.9 arcsec (frequencies 43 and/or 22 GHz) 
and for two other two sources we have 8.46-GHz observations with
resolution 2.3 arcsec. 
The highest-resolution observations available for 2238+00 are
from FIRST, and the source is unresolved.
In total four of the non-BAL QSOs are resolved:

0033$-$00 has core-lobe structure in the 8.46-GHz map. 
The total angular size is 6 arcsec (51 kpc) and 
the total spectral index is $\alpha_{4.86}^{8.46}=-0.55\pm0.09$.

A core-lobe structure is visible in 0125$-$00 at 4.86 and 8.46 GHz. 
Since at 4.86 GHz the lobe is not well resolved from the core, 
we only provide the total flux density (core and lobe) 
at this frequency in Table \ref{listfluxes2}. 
The angular size, as measured from the 8.46 GHz map, is 5 arcsec, 
corresponding to 42 kpc. 
The total spectral index of the source is $\alpha_{4.86}^{8.46}=-0.59\pm0.04$. 

1411+34 is resolved in the FIRST survey, and
our VLA maps at 4.86 and 8.46 GHz show
a core and double-lobe, with angular size at 8.46 GHz
of 23 arcsec (199 kpc). At 22 and 43 GHz only the core is detected.  
The spectral indices are  $\alpha_{4.86}^{8.46}=0.11\pm0.03$ for the core, 
$\alpha_{4.86}^{8.46}=-1.32\pm0.54$ for lobe A and $\alpha_{4.86}^{8.46}=-2.04\pm0.25$ for lobe B. 
The total spectral index is $\alpha_{4.86}^{8.46}=0.02 \pm 0.13$.

1728+56 is a double source at 4.86 and 8.46 GHz. 
At 22 GHz, a core component is also detected, and is coincident with
the optical position of the QSO.
The B and C components, which are resolved at 22 GHz, 
are blended at the lower frequencies. 
The angular size (separation of A and B) at 22 GHz is 10 arcsec, 
corresponding to 87 kpc. 
The spectral index calculation for the A component yields 
$\alpha_{4.86}^{8.46}=-1.04\pm0.04$ and for B+C $\alpha_{4.86}^{8.46}=-1.09\pm0.04$, 
indicating that the lobe emission dominates. For the total emission we found a spectral index 
$\alpha_{4.86}^{8.46}=-1.14\pm0.07$.\\

In summary, we found from our multi-wavelength observations
only eight resolved sources out of 59.
The fractions of resolved sources in the two samples are similar, as
are the ranges of angular sizes, from 20 to 200-400 kpc.
The morphologies of the BAL QSOs include one extended source,
two core-lobe and an ambiguous case between core-lobe
and core double-lobe.
The morphologies of the non-BAL QSOs include two core-lobe and
two core double-lobe sources.
The fraction of unresolved sources is 21/25=84\% for the BAL QSOs and 30/34=88\%
for the non-BAL QSOs. 20 of the unresolved BAL QSOs and 27 of the unresolved non-BAL QSOs
have been observed at the VLA at 8.46 GHz, with 2.3 arcsec resolution.
These data indicate that most of the QSOs in the two samples
have sizes below 20 kpc at 8 GHz, using the average redshift $z=2.4$ of
the two samples. Individual upper limits for the sources linear sizes are given in Table \ref{listfluxes1} and \ref{listfluxes2}. 
Two of the unresolved BAL QSOs, 1159+01 and 1624+37, were resolved by VLBA 
(Montenegro-Montes et al., 2008b, 2012 in preparation), both 
showing a core-jet morphology with sizes of 0.85 kpc and 60 pc respectively.

It has been suggested that up to 25\% of compact radio sources can
intrinsically be extended sources, viewed as being compact due to their
orientation (\citealt{Fanti}).  Our sample includes only
QSOs, which are usually considered to be active galactic nuclei
seen from a particular range of viewing angles, from a few degrees up to
$\sim$45$^\circ$ from the jet axis (limit imposed by the presence of
the dusty torus). 
This, in principle, could increase the contamination,
if we consider, as do Fanti et al., that a viewing angle
$<$20$^\circ$-30$^\circ$ can significantly reduce the projected linear size
of a source.

\begin{figure*}[tbp]
\centering
  \includegraphics[width=170mm]{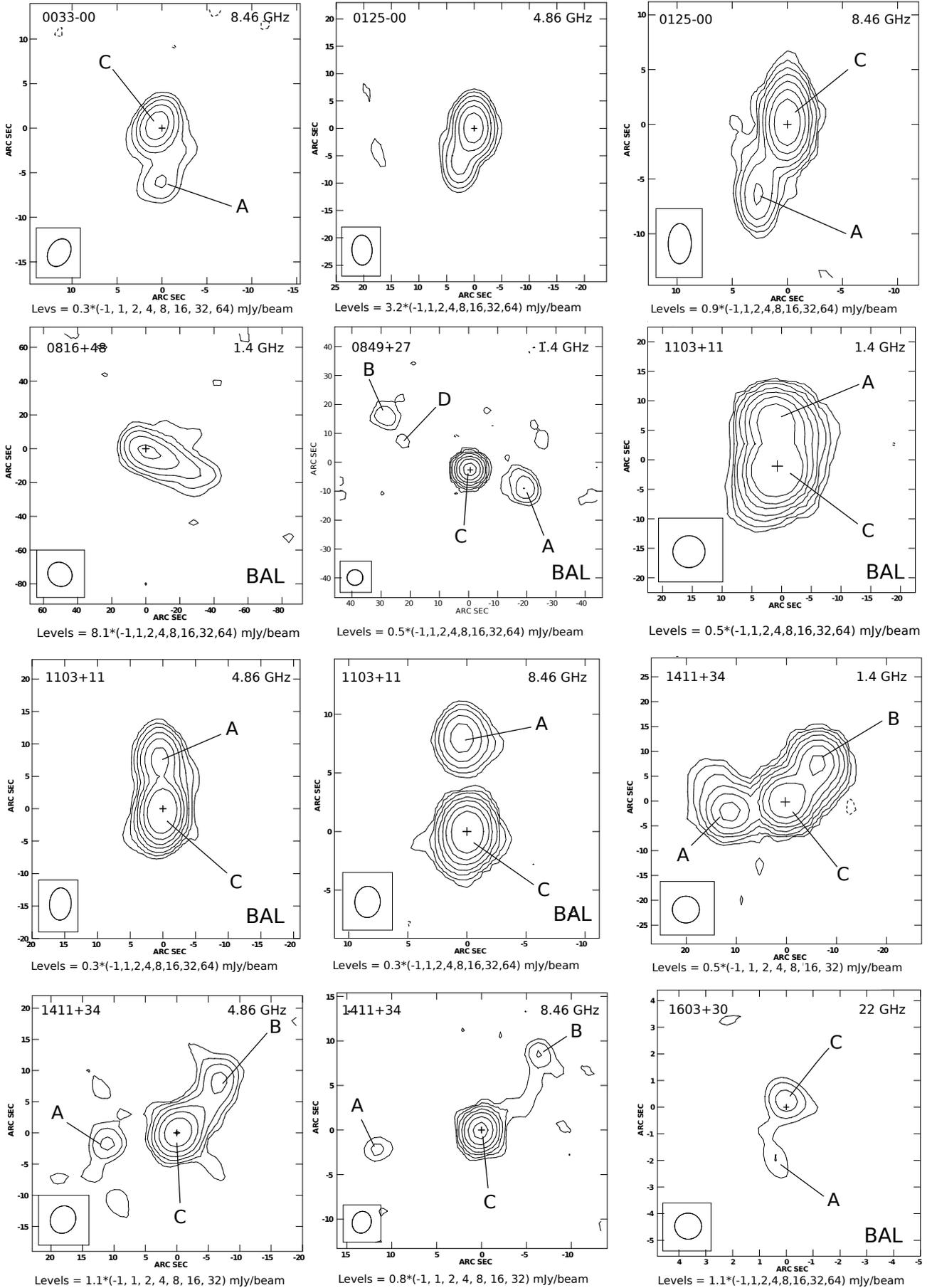}
\caption{Maps of the resolved QSOs. The synthesised beam size is shown 
in the lower left corner of the map. 
Levels are multiples of the 3-$\sigma$ flux density value in mJy/beam, according to the legend. 
A cross indicates the SDSS optical position.}
\label{morphology}
\end{figure*}

\begin{figure*}[tbp]
\centering
\addtocounter{figure}{-1}
  \includegraphics[width=170mm]{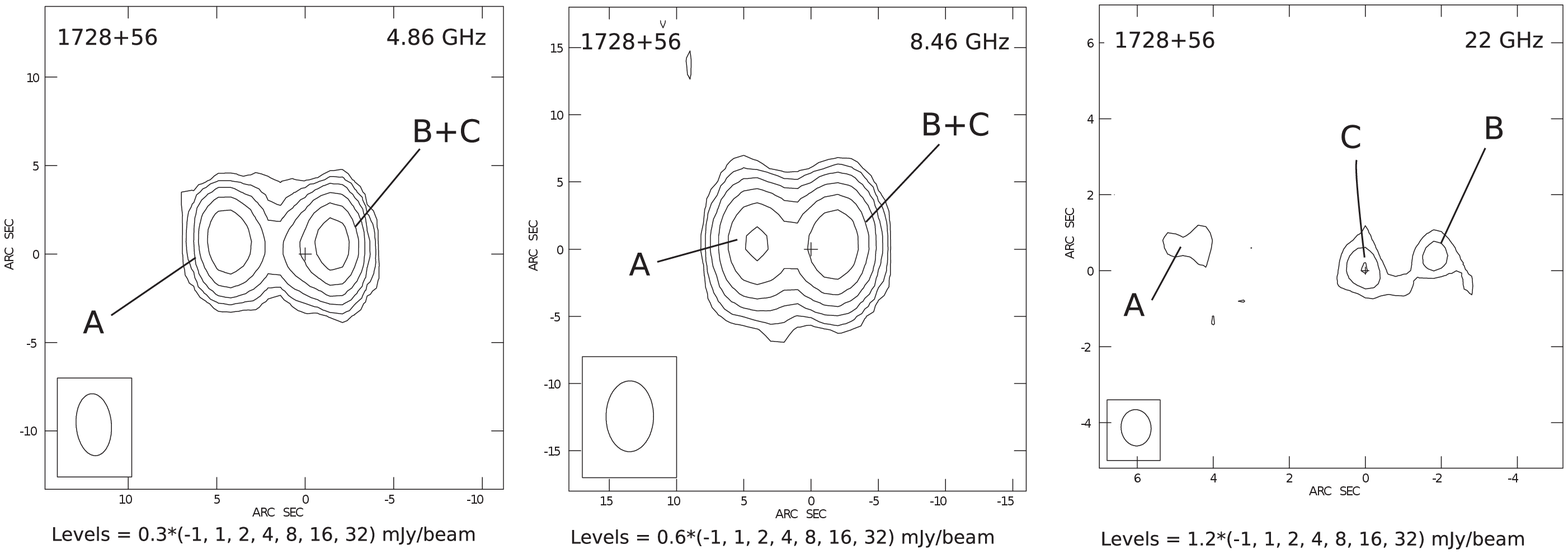}
\caption{Continued.}
\label{morphology}
\end{figure*}


\subsection{Variability}

For most of the sources we have two measurements of the flux
density at 4.8 GHz, one from Effelsberg observations
at 4.85 GHz and the other from the VLA at 4.86 GHz. 
Similarly, at 8.4 GHz, we have for many sources Effelsberg data at 8.35 GHz and
VLA data at 8.46 GHz. We checked these flux densities for potential
variable sources in the sample, evaluating for each pair
of measurements the fractional variability and significance. For 
the sources resolved in the VLA maps (see Sect. 4.1) we used the total flux densities. 
We adopted the fractional variability index defined by Torniainen et
al. (2005):

\begin{equation}
Var_{\Delta S} ={S_{max}  -  S_{min} \over  S_{min} }
\end{equation}

The significance of the source variability was estimated using
the $\sigma_{Var}$ parameter as defined e.g. by Zhou et al. (2006),

\begin{equation}
\sigma_{Var} = {|S_2 - S_1| \over \sqrt{\sigma_1^2 + \sigma_2^2}}
\end{equation}

where $S_i$ and $\sigma_i$ are the flux density and its corresponding
uncertainty. We consider as candidate variable sources those with $\sigma_{Var}\ge4$ and a fractional variability $\ge20\%$.
In the following we briefly discuss these cases. 

In the BAL QSO sample, 20 sources have 4.8-GHz 
flux densities from both Effelsberg and the VLA, and 17 have 
8.4-GHz flux densities (16 of these sources are in common with the first group).  
In four cases we found $\sigma_{Var} > 4$ and we list 
in Table \ref{BAL_var} the flux densities, variability significance, fractional variability 
and time interval. None of these fractional variabilities exceeds 20\%.

In the comparison sample, 30 sources were observed at
4.8 GHz by both telescopes, and 25 at 8.4 GHz.
Seven of the sources have variability significance above 4 at one or both frequencies and
they are listed in Table \ref{BAL_var}. 1005+48 shows
modest variability, just at the considered threshold.  0029$-$09 and 1521+43 show both a 
high variability significance, $\sigma_{Var} > 10$, and a high fractional
variability, $\sim$ 40-50 per cent, with 1521+43 being the most extreme case, 
showing large variations at the two frequencies. The remaining 
variable source, 1411+34, is resolved at the VLA (see Sect. 4.1). 
It shows variations at the two frequencies at a level 20-40$\%$, 
with significance 5-7$\sigma$.  
Since for this case we found the lower flux densities at the higher resolutions, the 
apparent variability for this source could be due to resolution effects and therefore it cannot be considered 
a bona fide intrinsic variable candidate.

Summarizing the results from the comparison between VLA and Effelsberg data, 
we found three sources with likely 
intrinsic variability, 0029$-$09, 1005+48 and 1521+43, all of them in the 
comparison sample. 
1411+34 (non-BAL QSO sample) shows flux-density variations that could be due to resolution effects. 
Given the small number of variable sources, 
it is not possible to firmly state whether BAL and non-BALs 
have different variability behaviour, 
although our data suggest a lower fraction of variables for the BAL QSO sample.

The BAL QSOs 1159+01, 1603+30 and 1624+37 were also included in the MM08
sample. We studied the possible variability of these sources
comparing the flux densities at various frequencies 
in this work with those reported at MM08, considering the same
radiotelescope (Effelsberg or VLA).  
Most of the flux densities for 1624+37 presented at MM08 were
taken from Benn et al. (2005).

For 1159+01 we found $\sigma_{var}> 4$ in the comparison of VLA data
at the frequencies of 8.4 GHz and 22 GHz, although in the first case
yielding a low fractional variability, of 9\%. At 22 GHz the flux
density variation is large, with $\sigma_{var}$=21 and fractional
variability 60\%. The variations could be due to 
resolution effects, since both correspond 
to an increase in flux density from the VLA A configuration data 
(HPBW=0.08 arcsec) from MM08, to the lower resolution 
VLA C configuration data (HPBW=0.9 arcsec) from this work. 
The flux densities varied from 160.8$\pm$1.25 mJy to 
176.6$\pm$2.2 at 8.4 GHz, and from 
105.5$\pm$1.15 mJy to 169.2$\pm$2.8 mJy at 22 GHz, 
in a time interval of 4.5 years.

For 1603+30 we could compare 2.6 and 4.8-GHz Effelsberg data as well
as 8.4 and 22-GHz VLA data. We found $\sigma_{var}$=4.9 and fractional
variability 49\% for the 4.8-GHz Effelsberg data, over an interval of 4.5 years,
indicating significant intrinsic variability.  For the 8.4-GHz 
VLA data there is flux density variation from $ 22.1 \pm 0.35$
mJy at VLA(A) to $26.9 \pm 0.6$ mJy for VLA(C), in 3.4 years. The variation is
significant, $\sigma_{var}=6.9$ and fractional variability 22\%,
although we cannot reject the possibility that resolution effects play a role. 
However, the variability at this frequency is confirmed by MM08, where 
the source is listed as a variable candidate (significance 
4.2 and fractional variability 22\%) from the comparison between the 
flux densities from their data and those from Becker et al. (2000), 
both from VLA in A configuration.

\begin{table*}
\centering
\caption{Sources with significant variability, 
$\sigma_{Var} > 4$, from the observations 
in this paper.  'BAL' in the last column indicates that the QSO
is from the BAL QSO sample.}
  \begin{tabular}{llcccccccccccc}
   \hline 
Name &&\multicolumn{4}{c}{4.8 GHz}& & \multicolumn{4}{c}{8.4 GHz} & &  Elapsed time &	\\
     &&   $S_{Eff}$ 	& $ S_{VLA}$ 	& $\sigma_{Var}$&  $Var_{\Delta S}$&	& $S_{Eff}$  	&  $S_{VLA}$  	& $\sigma_{Var}$   &   $Var_{\Delta S}$	& &\\ 
\hline
0029$-$09&~&      -        &  -            &  -   & -    &~ & 39.6$\pm$1.0 &  54.7$\pm$0.7 & 12.4~~ & 0.38 &~ &  21 d           &      \\
0756+37  &~& 209.6$\pm$2.6~~ & 226.2$\pm$2.0~~ &  5.1 & 0.08 &~ & -            &  -            & -    & -    &~ &  1.6 yr         & BAL  \\
1005+48  &~&  -            &  -            &  -   & -    &~ & 69.9$\pm$3.2 &  83.6$\pm$1.1 &  4.1 & 0.20 &~ &  1.6 yr         &	     \\
1103+11  &~&  -            &  -            &  -   & -    &~ & 65.9$\pm$1.1 &  75.0$\pm$0.8 &  6.7 & 0.14 &~ &  1.6 yr         & BAL  \\
1327+03  &~&  67.7$\pm$1.9 &  79.5$\pm$1.7 &  4.6 & 0.17 &~ & -            &  -            &  -   & -    &~ &  1.6 yr         & BAL  \\
1404+07  &~&  -            &  -            &  -   & -    &~ &173.6$\pm$2.4~~ & 189.6$\pm$2.2~~ &  4.9 & 0.09 &~ &  1.6 yr         & BAL  \\
1411+34  &~& 120.2$\pm$1.9~~ &  85.1$\pm$5.2 &  6.3 & 0.41 &~ &105.1$\pm$1.9~~ &  86.3$\pm$2.9 &  5.4 & 0.22 &~ &  1.6 yr, 1.6 yr &	     \\
1521+43  &~& 256.8$\pm$2.8~~ & 387.3$\pm$3.0~~ & 31.8~~ & 0.51 &~ &346.9$\pm$4.4~~ & 523.1$\pm$5.7~~ & 24.5~~ & 0.51 &~ &  1.6 yr, 1.6 yr &	     \\
1728+56  &~&  56.4$\pm$1.3 &  67.2$\pm$0.9 &  6.8 & 0.19 &~ & -            &   -           &  -   &  -   &~ &  1.6 yr         &	\\
2143+00  &~&  -            &   -           &  -   & -    &~ & 53.6$\pm$1.3 &  61.9$\pm$0.8 &  5.4 & 0.15 &~ &  21 d           &	\\
2331+01  &~&  -            &   -           &  -   & -    &~ & 19.3$\pm$0.6 &  23.0$\pm$0.5 &  4.7 & 0.19 &~ &  1.6 yr         &	\\
 \hline	
  \end{tabular}
    \label{BAL_var}
\end{table*}

For 1624+30, although the comparison of flux densities was 
possible at four frequencies (4.8 GHz from Effelsberg and 8.4, 10.5 and 22 
GHz from the VLA), none of them yielded $\sigma_{var}>4$. Therefore, from the 
comparison with the flux densities in MM08, only 1603+30 is a candidate 
variable.

In total four sources, one BAL QSO and three non-BAL QSOs, 
are classified as intrinsic variables (not due to resolution effects) 
at levels above 20\% (0029$-$09, 1005+48, 1521+43 and 1603+30). 
The additional data from MM08 weaken the trend 
of BAL QSOs being less variable than non-BAL QSOs.

Barvainis et al. (2005) studied the flux density variability at 8.4 GHz over 10 epochs 
(measurement interevals for each source ranging from two weeks to 1.6 years) 
of a core-dominated sample of 50 QSOs with 
$S_{\rm 8.4~ GHz} \ge 0.3$ mJy, including radio-quiet, 
radio-loud and radio-intermediate QSOs. 38 of the QSOs 
in their sample (76\%) have a flat radio spectrum, 
with $\alpha_{4.8}^{8.4} \ge -0.5$.  
The authors found five QSOs with fractional variability above 20\%, 
four in the range 20--40\% and one with fractional variability of 140\%.  
The four QSOs with higher variability have flat  
spectra, whereas the remaining one is steep. 
The fraction of sources varying by at least 20\%
found by Barvainis et al. (2005) (5/50) is consistent with 
our results for the SDSS-FIRST QSOs (4/59), within the errors. 
In addition, we will show in Section 4.4 that the variable sources 
in our work also tend to have a flat spectrum in the frequency 
range from 5 to 8 GHz, although our sample has 
a lower fraction of flat-spectrum sources (25/59 = 42\%, see Sect. 4.4 and 
Table \ref{listalpha1}, compared to 76\% in Barvainis et al. 2005 sample). 
This lower fraction of flat-spectrum sources could also explain the slightly lower 
fraction of variables in our work. 

Sadler et al. (2006) investigated the variability at 20 GHz 
over 1-2 years of a 
sample of radio sources selected to have
$S_{\rm 20~ GHz} \ge 100$ mJy and including 32 QSOs. 
The QSO sub-sample is dominated by flat-spectrum sources 
(69\% with $\alpha_5^8 \ge -0.5$) and has two sources 
just above the 20\% fractional variability threshold (2/32). 
Although the number of sources in this sample is small,
the proportion of variable sources is
consistent with that found by Barvainis et al. (2005) and in this work. 
We note however that the two QSOs in Sadler et al. (2006) with variability above 
20\% do not have flat spectra (their spectral indices being 
$\alpha_5^8=-0.59$ and $\alpha_5^8=-0.53$).


\subsection{Shape of the radio spectra}

\begin{figure*}[tbp!]
\centering
  \includegraphics[width=185mm]{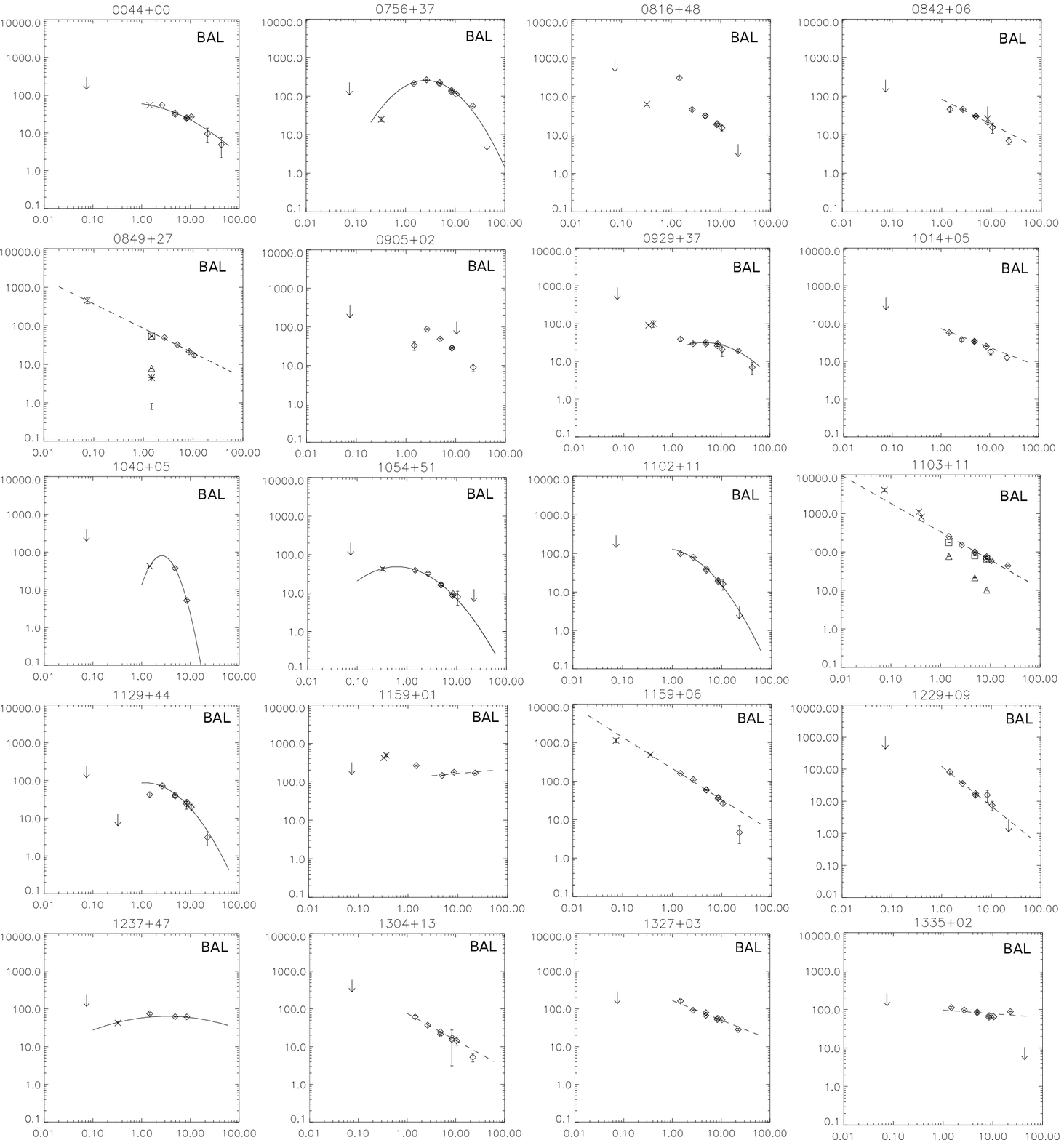}
\caption{Radio spectra of the 25 BAL and 34 non-BAL QSOs (GHz on
   x-axis, mJy on y-axis). Error bars are shown if larger than the
   symbol size. Crosses represent flux densities from the literature, rhombi 
   are flux densities from our observational campaign and arrows are
   upper limits from the literature. Solid lines are fits to a
   parabola and dashed lines are fits to a power-law. For
   resolved sources flux densities are presented as follows: a square
   for C component (core) and triangle, asterisk and dot for
   components A, B and D respectively (see Tables \ref{listfluxes1}
   and \ref{listfluxes2}). For 1728+56 the flux densities at 4.8 and
   8.4 GHz shown as asterisks correspond to the B+C components.}
\label{fits1} 
\end{figure*} 
 
\addtocounter{figure}{-1}
\begin{figure*}[tbp]
\centering
     \includegraphics[width=185mm]{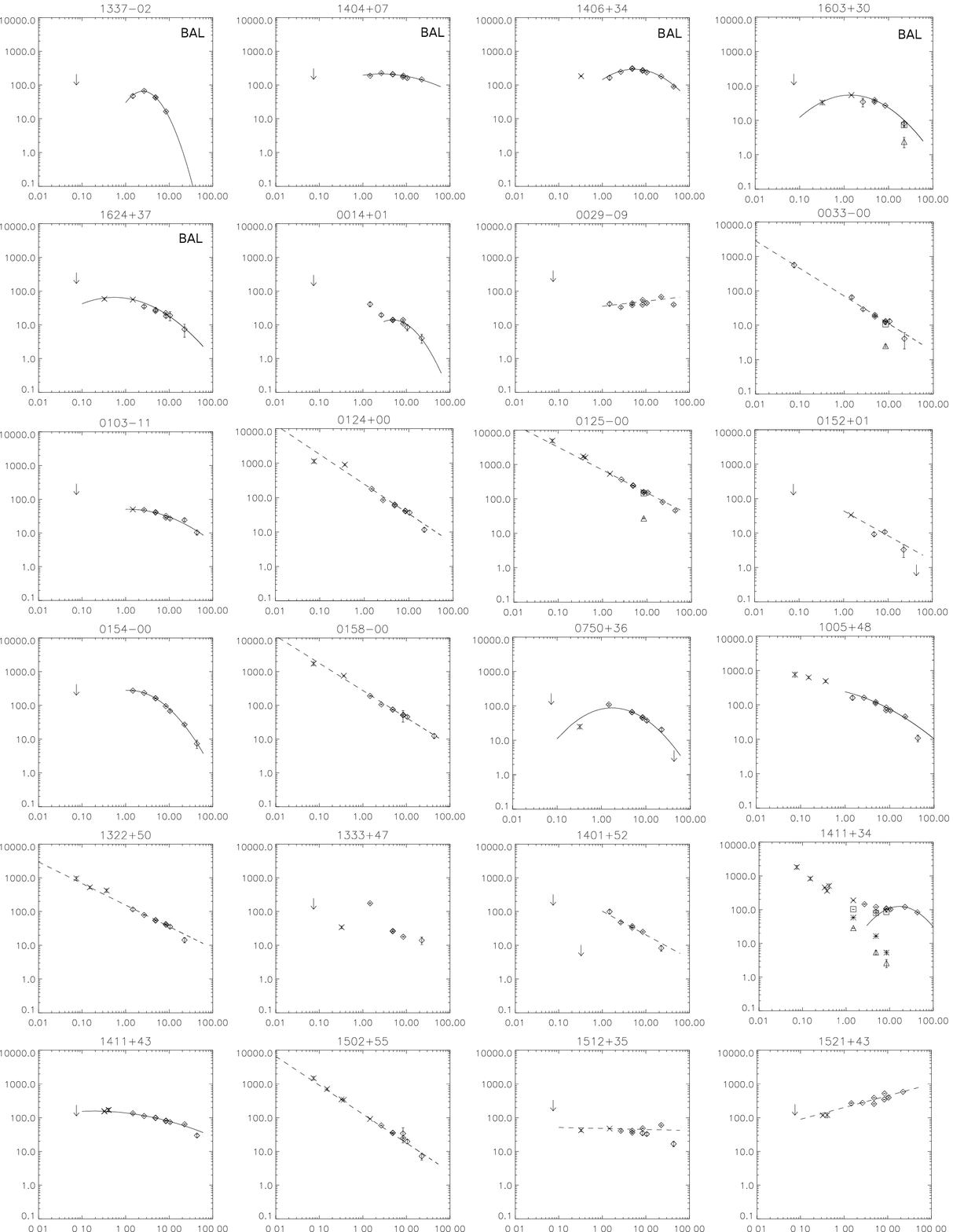}
  \caption{Continued.}\label{fits2}     
\end{figure*}

\addtocounter{figure}{-1}
\begin{figure*}[tbp]
\centering
     \includegraphics[width=185mm]{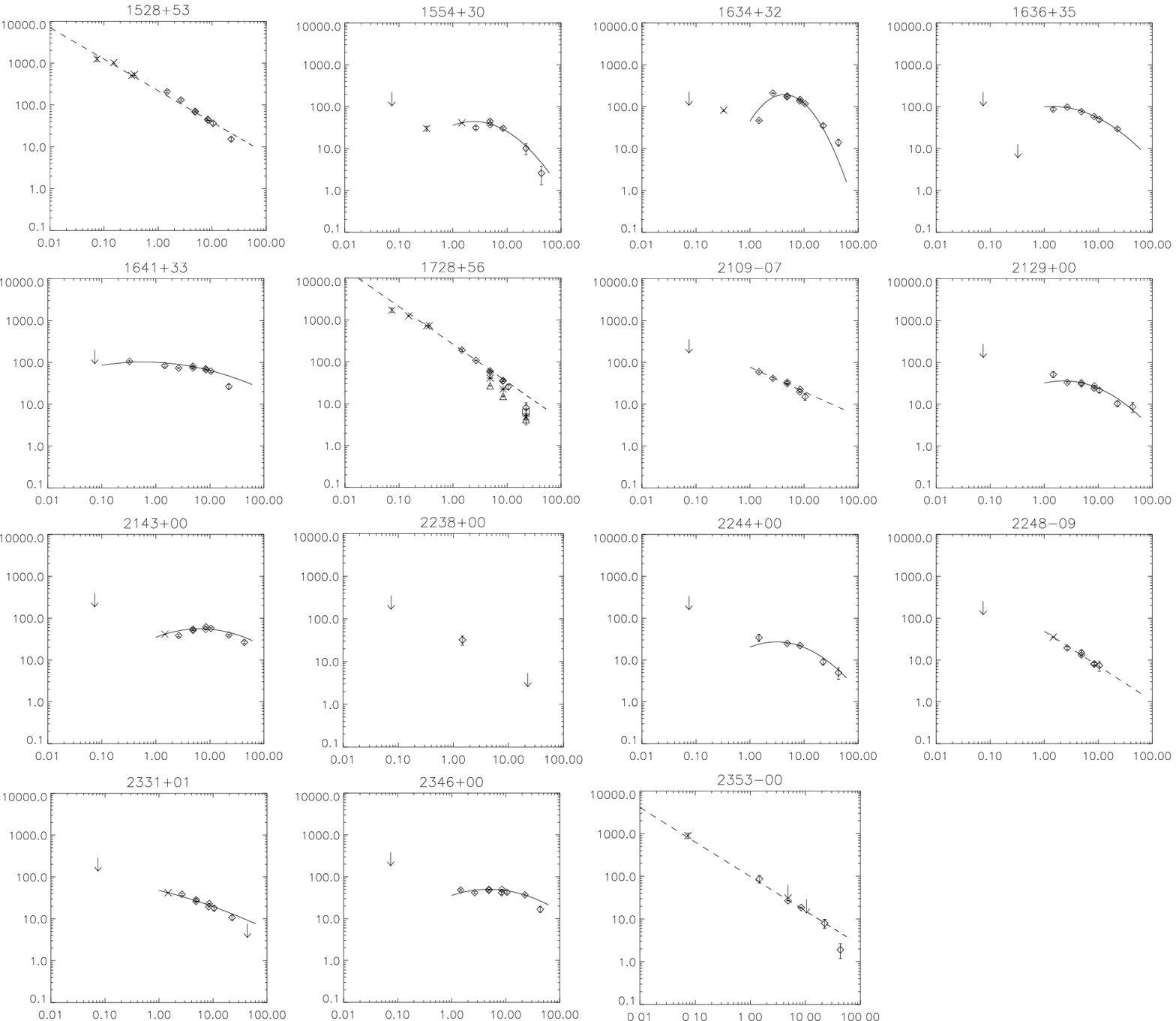}
  \caption{Continued.}\label{fits3}     
\end{figure*}

With the collected multi-frequency data it is possible to study the 
shape of the synchrotron emission of the quasars in the two samples,
allowing us to obtain the fraction of CSS-GPS sources.
GPS sources are compact ($\le$1 kpc) and have a convex radio spectrum
that peaks between 500 MHz and 10 GHz in the observer's frame, CSS
are larger (between 1 and 20 kpc in size) and have convex spectra
that tend to peak at lower frequencies, typically $<$500 GHz (\citealt{ODea}).  
The SEDs of the sources are shown in Figure 3 as log$S_{\nu}$ 
versus log$\nu$ plots, using the flux densities listed in Tables 
\ref{listfluxes1}, \ref{listfluxes2} and \ref{LF1}. 
For resolved sources we used the total flux densities.

The SEDs of the sources with observations at several
frequencies were fitted, via $\chi^2$ minimization,
with a power-law model (L) and a parabola (P),
on the log$S_{\nu}$ versus log$\nu$ plane.
A fit was accepted as statistically significant if the parameter Q, 
indicating the probability that a value of $\chi^2$ as poor as the value 
found should occur by chance, was above 0.01. 
The parabolic model was chosen as a simple representation for the SEDs 
curved on the log$S_{\nu}$ vs. log$\nu$ plane, following Kovalev (1996). 
For the cases where both models were statistically acceptable, the 
power-law was selected as the best fit. 
The fits are shown in Fig. 3 using a dashed line for the power-law model  
and a continuous line for the quadratic model.

For a total of 10 sources (BAL QSOs 0842+06, 
0849+27, 1014+05, 1229+09 and 1304+13, and non-BAL QSOs 
0033$-$00, 1401+52, 2109$-$07, 2248$-$09 and 2353$-$00) we found 
statistically significant fits as power laws, and these sources are 
labelled as ${\rm L_Q}$ in Table \ref{listalpha1}, 
where the fitting results are presented. 
For six further sources we found a statistically significant fit for 
the parabolic model 
(BAL QSOs 1054+51, 1102+11, 1337$-$02 and 1603+30, 
non-BAL QSOs  0154$-$00 and 1636+35). All these fits are convex, i.e. 
showing a flattening from high (10-20 GHz) to mid (1-5 GHz) 
frequencies. 
These sources are labelled as ${\rm P_Q}$ in Table \ref{listalpha1}.  
Although BAL QSO 1159+06 also falls in this category, we 
adopted for the source the linear model, since the peak of the parabola 
is far away from the observed frequencies, the linear 
and parabolic model being practically coincident over the observed 
range. We adopt the label $L$ in Table \ref{listalpha1} for this source.  

For the other 29 sources at least one of the models shows a good match 
from visual inspection, and we selected as the best model the one 
yielding the lowest mean squared error, labelling the sources as L or 
P in Table \ref{listalpha1}. The linear model includes six sources (BAL QSO 
1327+03 and non-BAL QSOs 0029$-$09, 0152+01, 0158$-$00, 1322+50 and 1512+35), and 
the parabolic model 15 sources (BAL QSOs 0044+00, 0756+37, 1129+44, 
1237+47, 1404+07 and 1624+37 and non-BAL QSOs 0103$-$111, 0750+36, 
1411+43, 1641+33, 2129+00, 2143+00, 2244+00, 2331+01 and 2346+00), all 
of them with convex shape. Although the remaining eight sources have 
the lowest mean squared error for the quadratic model, we adopted the 
linear fit, also with a low mean squared error, because either the 
peak was far away from the SED, making the linear and quadratic model 
very similar, or the parabolic shape was concave, which is 
inconsistent with the expected shapes from synchrotron models.  As for 
1159+06, described in the previous paragraph, we used the italic label $L$ 
for these sources (BAL QSOs 1103+11 and 1335+02, and non-BAL QSOs 0124+00, 0125$-$00, 
1502+55, 1521+43, 1528+53 and 1728+56). For the sources modelled with a parabola, with its peak within 
the fitted range, the frequency peaks are listed in Table 9.

Another three sources were not fitted, since they showed abrupt 
changes in their SEDs (BAL QSOs 0816+48 and 0905+02, 
and non-BAL QSO 1333+47).

Another eight sources (BAL QSOs 0929+37, 1159+01, 1406+34, and non-BAL QSOs 0014+01, 1005+48, 1411+34, 1554+30, 1634+32) 
have SEDs that suggest the presence of a separate component at low 
frequency. 
In fact 1411+34 was also morphologically resolved 
as a core double-lobe with the two lobes steeper, i.e. stronger at 
low frequencies, than the core (see Sect. 4.1). 
For these sources we considered fits removing one or various 
of the lowest-frequency data points, with hints of excess emission. 
The low-frequency points rejected for the fits are indicated in 
Table \ref{listalpha1}.
Regarding the high-frequency components, six of the sources 
belong to class P and another one to class $\rm P_Q$, all with convex shape.
The SED of 1159+01 shows hints of excess 
emission from 325 MHz to 1.4 GHz, leaving only three high-frequency 
data points. Since the power law fit gives a low mean squared error 
we adopted this model for the source.  
For the cases where the high-frequency component was modelled as a 
parabola with its peak within the fitted range, the frequency peaks 
are listed in Table \ref{listalpha1}, using italic digits. 
However, these sources 
cannot be considered as CSS-GPS candidates, because 
of the presence of the secondary low-frequency emission.

The remaining sources in the sample are 1040+05 (BAL QSO) and 2238+00.
For 1040+05 only three data points are available, with an obvious
flattening at low frequencies, and we selected as best model the
parabola passing through these points.  2238+00 has a good quality
measurement only at one frequency.\\

In total we found nine BAL QSOs and 16 non-BAL QSOs whose complete SEDs are
consistent with power laws.  For 11 BAL QSOs and 11 non-BAL QSOs the fits
indicate a curved shape along the whole SED due to a flattening of the spectra at low
frequencies.  For 15 of these (eight BAL QSOs and seven non-BAL QSOs) 
the frequency peak of the model parabola falls within the 
fitted range, with values ranging from 500 MHz to 7 GHz in the 
observer frame, indicating that they represent candidate GPS sources. 
Three sources that were not fitted due to abrupt changes in their SEDs have
a maximum within the observed frequency range 
with frequency peaks above 1 GHz. 
In addition, although BAL QSO 1129+44 is fitted with a parabola with its peak 
below the observed frequency range, the flux-density distribution shows a peak 
at 2.6 GHz. 
Among these 19 sources whose SED is GHz-peaked, two are resolved,
BAL QSOs 0816+48 and 1603+30, with sizes of 217 kpc and 17 kpc respectively (see Section 4.1),
exceeding the limit of 1 kpc for GPS sources (\citealt{ODea}). Excluding these two sources
the total number of candidate GPS sources
would be 9 BAL QSOs and 8 non-BAL QSOs, with corresponding fractions with respect to 
the total samples of $36\pm 12$\% (9/25) and $23 \pm 8$\% (8/34), adopting Poisson errors. 
In Section 4.1 we obtained for the unresolved sources a 
conservative upper limit of 20 kpc for their sizes 
at 8.4 GHz,
therefore higher-resolution observations are needed to confirm the GPS classification. 
In particular, 
this classification is confirmed for 1624+37, with a size of 60 pc at 5 GHz and 75 pc at 8 GHz 
(Montenegro-Montes et al. 2008b, 2012 in preparation). 
The fractions of GPS candidates in the BAL QSO and non-BAL QSO samples are similar, 
within the errors. Considering the interpretation that GPS sources
are young, our result suggests that BAL QSOs are not a 
younger population than non-BAL QSOs.

We adopted a conservative approach in the quantification of young objects,
since only candidate GPS sources were considered for that purpose, but also CSS objects,
showing a steep spectrum in the GHz frequency range, with peak frequencies 
below 500 MHz, can be interpreted as young sources. 
Additional observations at low frequency could confirm the presence
of peaks in the MHz frequency range for some of the sources in this work. 
In fact the maximum size for CSS is 20 kpc (\citealt{ODea}) 
and most of the sources in this work are unresolved with sizes below this limit.

 

The fraction of QSOs with hints of an additional low-frequency
component (up-turn at low frequency) 
is $12 \pm 7$\% for BAL QSOs (3/25) and $15 \pm 7$\% for non-BAL QSOs
(5/34), the two values being similar within the errors. Since this 
low-frequency excess emission likely corresponds to old components, 
this result again favours similar ages for BAL and non-BAL QSOs. 



\begin{table*}
\centering
\caption{Radio spectral shape and spectral indices of the QSO sample}
\label{listalpha1}
\begin{tabular}{lcrrcrrcc}
\hline
\multicolumn{1}{c}{Name}  			&  
\multicolumn{1}{c}{Fit type} 			&
\multicolumn{1}{c}{$\nu_{peak}$}  		&  
\multicolumn{1}{c}{$\nu_{peak}^{rest}$}		&
\multicolumn{1}{c}{$\nu_{rej}$}			&
\multicolumn{1}{c}{$\alpha_{4.8}^{8.4}$} 	&
\multicolumn{1}{c}{$\alpha_{8.4}^{22}$} 	& 
\multicolumn{1}{c}{  log10($L_{\rm 4.8~GHz}$)}  \\
\multicolumn{1}{c}{         }  &  
\multicolumn{1}{c}{         }  &  
\multicolumn{1}{c}{(GHz)    }  &  
\multicolumn{1}{c}{(GHz)    }  &  
\multicolumn{1}{c}{(GHz)    }  &  
\multicolumn{1}{c}{         }  &  
\multicolumn{1}{c}{         }  & 
\multicolumn{1}{c}{(W Hz$^{-1}$)}  \\  
\multicolumn{1}{c}{(1)} &
\multicolumn{1}{c}{(2)} &
\multicolumn{1}{c}{(3)} &
\multicolumn{1}{c}{(4)} &
\multicolumn{1}{c}{(5)} &
\multicolumn{1}{c}{(6)} &
\multicolumn{1}{c}{(7)} &
\multicolumn{1}{c}{(8)} &
\multicolumn{1}{c}{(9)} \\
%
\hline
          &             &              &              &                   &                   &                      &       &    \\
0044+00	  & P	        &- 	       & -	      &             -     &  $-$0.53$\pm$0.10 &   $-$1.01$\pm$0.29   & 26.89 &BAL \\	  
0756+37	  & P	        &2.5	       &8.7	      &             -     &  $-$0.84$\pm$0.03 &   $-$0.95$\pm$0.05   & 27.97 &BAL \\      
0816+48	  & -	        &-	       & -	      &             -     &  $-$0.86$\pm$0.05 &   $\le-$1.26~~           & 27.47 &BAL \\   	  
0842+06	  & L$_{\rm Q}$ &-	       & -	      &             -     &  $-$0.62$\pm$0.05 &   $-$1.13$\pm$0.21   & 26.94 &BAL \\   	  
0849+27	  & L$_{\rm Q}$ &-	       & -	      &             -     &  $-$0.73$\pm$0.12 & 	-~~~~~~~     & 26.69 &BAL \\	  
0905+02	  & -	        &-	       & -	      &             -     &  $-$0.93$\pm$0.13 &   $-$1.20$\pm$0.23   & 27.01 &BAL \\   	  
0929+37	  & P$_{\rm Q}$ &\textit{4.5}  &\textit{13.1} &0.325, 0.408, 1.4  &  $-$0.16$\pm$0.04 &   $-$0.44$\pm$0.08   & 26.53 &BAL \\	  
1014+05	  & L$_{\rm Q}$ &-   	       &-	      &             -     &  $-$0.57$\pm$0.05 &   $-$0.70$\pm$0.18   & 26.80 &BAL \\	  
1040+05	  & P$_{\rm Q}$ &2.6	       &8.8	      &             -     &  $-$3.55$\pm$0.15 & 	-~~~~~~~     & 28.61 &BAL \\ 	  
1054+51	  & P$_{\rm Q}$ &0.6	       &2.0	      &             -     &  $-$0.93$\pm$0.13 &   $\le$ 0.29~~~~         & 26.79 &BAL \\	  
1102+11	  & P$_{\rm Q}$ &-	       &-	      &             -     &  $-$1.25$\pm$0.07 &   $\le-$1.59~~         & 27.36 &BAL \\  	  
1103+11	  & $L$         &-	       &-	      &             -     &  $-$0.53$\pm$0.02 &   $-$0.55$\pm$0.04   & 27.11 &BAL \\       
1129+44	  & P	        &-	       &-	      &             -     &  $-$0.67$\pm$0.16 &   $-$2.20$\pm$0.40   & 26.98 &BAL \\ 	  
1159+01	  & L	        &-             &-	      &0.325, 0.365, 1.4  &     0.34$\pm$0.03 &   $-$0.04$\pm$0.02   & 26.98 &BAL \\	  
1159+06	  & $L$	        &-   	       &-	      &             -     &  $-$0.80$\pm$0.05 &   $-$2.15$\pm$0.51   & 27.22 &BAL \\	  
1229+09   & L$_{\rm Q}$ &-	       &-	      &             -     &  $-$0.17$\pm$0.73 &   $\le-$1.80~~         & 26.53 &BAL \\	  
1237+47	  & P	        &3.4           &11.1	      &             -     &  $-$0.02$\pm$0.05 & 	-~~~~~~~     & 26.88 &BAL \\ 	  
1304+13	  & L$_{\rm Q}$ &-	       &-	      &             -     &  $-$0.66$\pm$0.06 &   $-$1.20$\pm$0.25   & 26.93 &BAL \\	  
1327+03	  & L	        &-   	       &-	      &             -     &  $-$0.62$\pm$0.04 &   $-$0.70$\pm$0.07   & 27.50 &BAL \\	  
1335+02	  & $L$	        &-             &-	      &             -     &  $-$0.40$\pm$0.04 &      0.25$\pm$0.02   & 27.04 &BAL \\	  
1337$-$02 & P$_{\rm Q}$ &2.5	       &10.2	      &             -     &  $-$1.73$\pm$0.06 & 	-~~~~~~~     & 27.99 &BAL \\      
1404+07	  & P           &2.6	       &10.1	      &             -     &  $-$0.15$\pm$0.03 &   $-$0.26$\pm$0.02   & 27.66 &BAL \\	  
1406+34	  & P	        &\textit{5.4}  &\textit{19.2} &0.325              &  $-$0.22$\pm$0.03 &   $-$0.44$\pm$0.02   & 27.79 &BAL \\	  
1603+30	  & P$_{\rm Q}$ &1.4	       &4.2	      &             -     &  $-$0.45$\pm$0.13 &   $-$1.04$\pm$0.14   & 26.75 &BAL \\   	  
1624+37	  & P           &0.5	       &2.2	      &             -     &  $-$0.78$\pm$0.29 &   $-$0.92$\pm$0.44   & 27.32 &BAL \\     
          &             &              &              &                   &                   &                      &       &    \\
\hline
          &             &              &              &                   &                   &                      &       &    \\
0014+01	  & P	        &\textit{4.8}  &\textit{15.3} &1.4, 2.6           &  $-$0.46$\pm$0.09 &   $-$1.02$\pm$0.31   & 26.43 &    \\	    
0029$-$09 & L	        &  -           & -    	      & -                 &     0.63$\pm$0.06 &      0.22$\pm$0.04   & 26.44 &    \\	    
0033$-$00 & L$_{\rm Q}$ &  -           & -    	      & -                 &  $-$0.55$\pm$0.09 &   $-$1.19$\pm$0.50   & 26.39 &    \\   	    
0103$-$11 & P	        &  -           & -            & -                 &  $-$0.41$\pm$0.04 &   $-$0.29$\pm$0.09   & 26.87 &    \\      	   
0124+00	  & $L$         &  -           & -            & -                 &  $-$0.80$\pm$0.06 &   $-$1.27$\pm$0.14   & 27.09 &    \\	    
0125$-$00 & $L$	        &  -           & -            & -                 &  $-$0.59$\pm$0.04 &   $-$0.78$\pm$0.03   & 27.77 &    \\         
0152+01	  & L & -  & - & -                &     0.27$\pm$0.28 &   $-$1.20$\pm$0.41   & 26.12 &    \\            
0154$-$00 & P$_{\rm Q}$ &  -           & -            & -                 &  $-$0.97$\pm$0.06 &   $-$1.28$\pm$0.07   & 27.56 &    \\         
0158$-$00 & L	        &  -           & -            & -                 &  $-$0.64$\pm$0.03 &	    -~~~~~~~	     & 27.42 &    \\	    
0750+36	  & P	        &  1.7         & 5.1          & -                 &  $-$0.60$\pm$0.03 &   $-$0.85$\pm$0.11   & 27.10 &    \\	    
1005+48	  & P	        &  -           & -            &0.074, 0.151, 0.365&  $-$0.67$\pm$0.03 &   $-$0.62$\pm$0.04   & 27.55 &    \\	    
1322+50	  & L	        &  -           & -            & -                 &  $-$0.58$\pm$0.04 &   $-$1.06$\pm$0.19   & 26.87 &    \\	    
1333+47	  &	-       &  -           & -            & -                 &  $-$0.68$\pm$0.06 &   $-$0.25$\pm$0.26   & 26.98 &    \\    	   
1401+52	  & L$_{\rm Q}$ &  -           & -            & -                 &  $-$0.52$\pm$0.09 &   $-$1.12$\pm$0.20   & 27.11 &    \\	    
1411+34	  & P	        &\textit{16.8} &\textit{47.4} &0.074 - 4.86       &     0.02$\pm$0.13 &      0.34$\pm$0.04   & 26.83 &    \\	    
1411+43	  & P	        &  -           & -     	      & -                 &  $-$0.35$\pm$0.03 &   $-$0.26$\pm$0.03   & 27.55 &    \\	    
1502+55	  & \textit{L}  &  -           & -            & -                 &  $-$0.64$\pm$0.05 &   $-$1.27$\pm$0.24   & 27.31 &    \\	    
1512+35	  & L 	        &  -           & -            & -                 &     0.30$\pm$0.04 &      0.23$\pm$0.05   & 26.52 &    \\      	   
1521+43	  & \textit{L}  &  -           & -    	      & -                 &     0.54$\pm$0.02 &      0.11$\pm$0.02   & 27.36 &    \\  	    
1528+53	  & $L$         &  -           & -            & -                 &  $-$0.89$\pm$0.10 &   $-$1.07$\pm$0.10   & 27.61 &    \\          
1554+30	  & P	        &\textit{2.4}  &\textit{8.8}  &0.325              &  $-$0.36$\pm$0.16 &   $-$1.12$\pm$0.30   & 26.98 &    \\          
1634+32	  & P	        &\textit{4.3}  &\textit{14.4} &0.325              &  $-$0.34$\pm$0.03 &   $-$1.47$\pm$0.10   & 27.53 &    \\	
1636+35	  & P$_{\rm Q}$ &  1.4         & 4.1          & -                 &  $-$0.49$\pm$0.11 &   $-$0.70$\pm$0.09   & 27.05 &    \\     	   
1641+33	  & P           &  0.6         & 2.2          & -                 &  $-$0.38$\pm$0.03 &   $-$0.93$\pm$0.13   & 27.35 &    \\	    
1728+56	  & $L$         &  -           & -    	      & -                 &  $-$1.14$\pm$0.07 &   $-$0.86$\pm$0.11   & 27.22 &    \\          
2109$-$07 & L$_{\rm Q}$ &  -           & -            & -                 &  $-$0.56$\pm$0.06 &	    -~~~~~~~	     & 26.68 &    \\  	    
2129+00	  & P	        &  2.3         & 9.1          & -                 &  $-$0.32$\pm$0.04 &   $-$1.00$\pm$0.16   & 26.99 &    \\	    
2143+00	  & P	        &  6.6         & 20.1         & -                 &     0.35$\pm$0.03 &   $-$0.46$\pm$0.05   & 26.53 &    \\	    
2244+00	  & P	        &  3.1         & 12.2	      & -                 &  $-$0.21$\pm$0.06 &   $-$0.93$\pm$0.17   & 26.79 &    \\	    
2248$-$09 & L$_{\rm Q}$ &  -           & -            & -                 &  $-$0.88$\pm$0.12 &	    -~~~~~~~	     & 26.58 &    \\	    
2331+01	  & P	        &  -           & -            & -                 &  $-$0.21$\pm$0.05 &   $-$0.78$\pm$0.14   & 26.73 &    \\	    
2346+00	  & P	        &  4.9         & 13.7	      & -                 &     0.01$\pm$0.04 &   $-$0.31$\pm$0.05   & 26.58 &    \\	    
2353$-$00 & L$_{\rm Q}$ &  -           & -    	      & -                 &  $-$0.67$\pm$0.06 &   $-$0.87$\pm$0.23   & 26.68 &    \\            
\hline
\end{tabular}
\begin{list}{}{}
\item[{\bf Notes:}] 
Column 2:  Best-fitting function is specified 
(L for linear, P for parabolic, see Sect. 4.3); 
cols. 3 and 4: Peak frequencies (observer and rest-frame); 
col. 5: Frequencies of the rejected  flux densities for the fits (if any); 
cols 6-7: Observer-frame spectral indices; 
col. 8: Radio luminosity; 
col. 9: Indication whether the source is a BAL QSO.
\end{list}

\end{table*}


\subsection{Spectral indices}

\begin{figure}[tbp]
\centering 
\includegraphics[width=100mm]{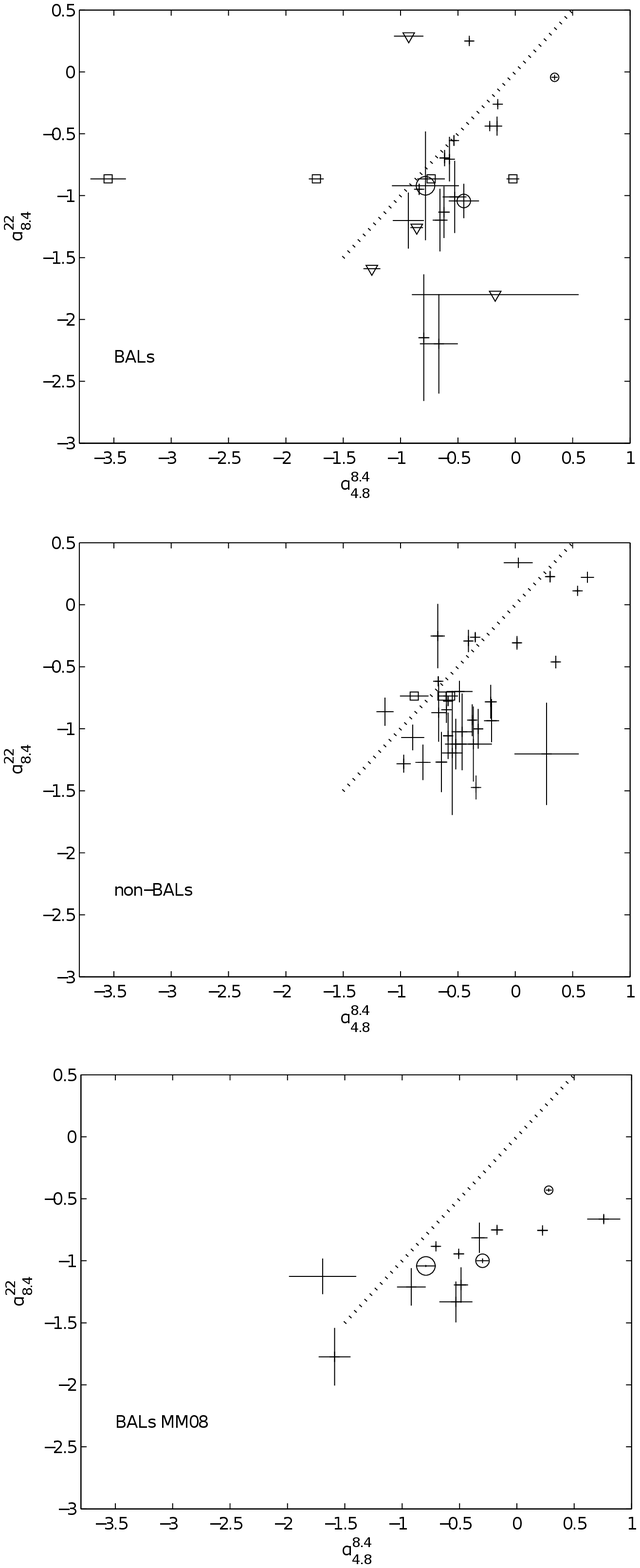}
\caption{$\alpha_{8.4}^{22}$ versus $\alpha_{4.8}^{8.4}$ for the BAL QSOs, 
the non-BAL QSOs and MM08 sample. 
The dashed line traces the locus of power-law spectra.
Square symbols indicate sources lacking a 22-GHz flux density. 
For these, we adopted the $\alpha_{8.4}^{22}$ mean value of the sample. 
Upper limits for $\alpha_{8.4}^{22}$ are plot as triangles.
Circles indicate objects in common with MM08 
(symbol size increases with the RA of the source).}
\label{alphas}	
\end{figure}

The distribution of radio spectral indices 
constrains the distribution of orientations for a given population
of radio sources (Orr \& Browne 1982), since flatter spectral indices 
imply lines of sight closer to the radio axis.
Spectral indices of the QSOs were computed in the observed 
frequency intervals 4.8-8.4 GHz and 8.4-22 GHz, since these 
frequencies exceed the typical peak frequencies of the 
candidate GPS sources in the sample.  
We used the total flux densities for the resolved sources 
and VLA data, which were obtained during a one week run 
(see Table 3). 
Effelsberg flux densities were adopted for the few
sources/frequencies lacking VLA data 
(BAL QSOs 0849+27 and 1229+09 and
non-BAL QSOs 0154$-$00 and 1636+35). 
%
%
The spectral indices and their errors
are listed in Table \ref{listalpha1}.

Figure \ref{alphas} (top and middle panels) 
show $\alpha_{8.4}^{22}$ versus $\alpha_{4.8}^{8.4}$
for the two samples.  
The square symbols correspond to sources without
available spectral index $\alpha_{8.4}^{22}$.
They were plotted using the mean
$\alpha_{8.4}^{22}$ of each sample. 
Upper limits are plotted
as triangles.
The dotted line traces pure power-laws: 
the location of most of the sources below this line 
is due to the steepening at high frequencies. 
The spectral indices of the BAL QSOs show a large scatter 
in the plot.  
Although most BAL QSOs are found in the same region as non-BAL 
QSOs, there is an apparent excess of steep sources in the 
BAL QSO sample, with six sources showing 
spectral indices below $-1.5$ at either of the two 
frequency ranges.
In addition, there appears to be an excess of non-BAL QSO 
sources with $\alpha_{4.8}^{8.4} > 0$. 
However, when the distributions of spectral indices 
for the BAL and non-BAL QSO samples are compared using
the Kolmogorov-Smirnov (K-S) test, 
they are found to differ only at a significance level of 74\% for
$\alpha_{4.8}^{8.4}$ and 0.3\% for $\alpha_{8.4}^{22}$. 
Considering the test that the distribution of spectral indices 
is steeper for BAL QSOs than for non-BAL QSOs, the significance 
levels increase to 87\% for
$\alpha_{4.8}^{8.4}$ and to 28\% for $\alpha_{8.4}^{22}$.
It is also useful to test the hypothesis that BAL QSOs 
are flatter than non-BAL QSOs. This hypothesis can be 
rejected at a 100\% confidence level for 
$\alpha_{4.8}^{8.4}$ and at 77\% confidence level for 
$\alpha_{8.4}^{22}$. 

\begin{figure*}[tbp]
\centering
\includegraphics[width=180mm]{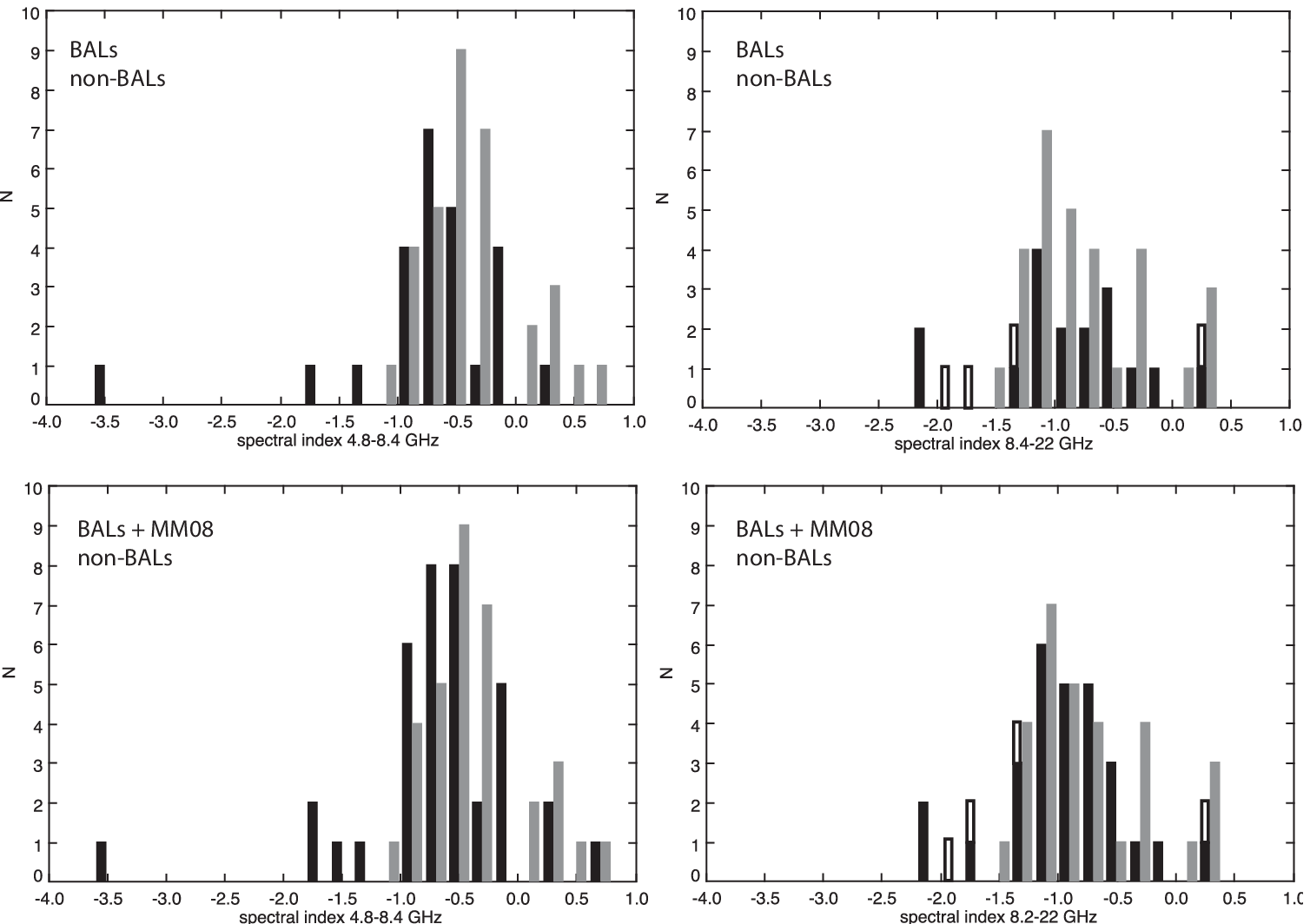}
\caption{Distribution of the radio spectral indices ($\alpha_{4.8}^{8.4}$ and $\alpha_{8.4}^{22}$) for the 
BAL (black) and non-BAL (grey) QSO samples in this work. Upper limits are in white.}
\label{KS}	
\end{figure*}

\begin{table*}
\centering
\caption{Spectral-index statistics for various QSO samples: the BAL and 
non-BAL QSO samples presented in this paper (`BAL', `non-BAL'), 
the BAL QSO sample presented in MM08, and the comparison QSO sample from Vigotti et al. 1997, 1999.} 
\label{spec_stat}
\begin{tabular}{lcccccccccccccc}
\hline
& & & & & & & & & & & & & &\\
\multicolumn{1}{c}{}  				&  
\multicolumn{6}{c}{$\alpha_{4.8}^{8.4}$}  	& 
\multicolumn{1}{c}{}  				& 
\multicolumn{1}{c}{}  				&
\multicolumn{6}{c}{$\alpha_{8.4}^{22}$}  	\\
\multicolumn{1}{c}{}  				&
\multicolumn{1}{c}{N}  				&
\multicolumn{1}{c}{min}  			&
\multicolumn{1}{c}{max}  			&
\multicolumn{1}{c}{median}  			&
\multicolumn{1}{c}{mean}  			&
\multicolumn{1}{c}{std}  			&
\multicolumn{1}{c}{}  				&
\multicolumn{1}{c}{}  				&
\multicolumn{1}{c}{N}  				&
\multicolumn{1}{c}{min}  			&
\multicolumn{1}{c}{max}  			&
\multicolumn{1}{c}{median}  			&
\multicolumn{1}{c}{mean}  			&
\multicolumn{1}{c}{std}  			\\
\hline
BAL      		   & 25  &  $-$3.55 & 0.34 &  $-$0.62 &$-$0.71 & 0.72 & & & 20 & $-$2.20 & 0.25 & $-$0.92 & $-$0.86 &  0.64 \\
BAL+MM08 		   & 37  &  $-$3.55 & 0.76 &  $-$0.62 &$-$0.66 & 0.70 & & & 32 & $-$2.20 & 0.25 & $-$0.93 & $-$0.93 &  0.54 \\
non-BAL  		   & 33  &  $-$1.14 & 0.63 &  $-$0.49 &$-$0.39 & 0.43 & & & 30 & $-$1.47 & 0.34 & $-$0.87 & $-$0.74 &  0.50 \\
B3-VLA QSOs$^1$ ($<$2.3'') & 40  &  $-$1.38 & 0.54 &  $-$0.65 &$-$0.57 & 0.47 & & & -  &   -     & -    &   -     &   -     &   -   \\
B3-VLA QSOs$^1$ (all)           & 123 &  $-$2.23 & 0.54 &  $-$0.89 &$-$0.79 & 0.45 & & & -  &   -     & -    &   -     &   -     &   -   \\
\hline
\end{tabular}

\begin{list}{}{}
\item[$^{\mathrm{1}}$] The spectral index for this sample is $S_{4.8}^{10.6}$.
\end{list}

\end{table*}

Although we do not find statistical evidence of steeper spectra 
for BAL QSOs, we can firmly reject that BAL QSOs have on average 
flatter radio spectra than non-BAL QSOs, 
in the frequency range 4.8-8.4 GHz.

Figure 4 (bottom panel) shows the same spectral index 
diagram for the BAL QSOs in MM08. 
BAL QSOs 1159+01, 1603+30 and 1624+37, which are common 
to both samples, were plotted with a different symbol
(circles with their size increasing with the right ascension 
of the sources). 
From the combined sample of BAL QSOs from this work and MM08 
(using our data for the sources in common), we find that 
the hypothesis that BAL QSOs have steeper 
$\alpha_{4.8}^{8.4}$ than non-BAL QSOs has a higher 
confidence level, of 91\%, although still below the threshold 
of 95\% generally adopted for the rejection of the 
null hypothesis. 
It is important to note that in this test 
we are using the comparison sample selected for this work, 
with $S_{1.4} > 30$ mJy, 
brighter than the 15-mJy limit of MM08 .
Regarding $\alpha_{8.4}^{22}$, the combined sample of BAL QSOs 
has steeper spectra than the non-BAL QSO sample only at a 
55\% confidence level. 
The hypothesis that the BAL QSOs in the combined sample 
have flatter spectra than non-BAL QSOs can be rejected at a 
97\% confidence level 
for $\alpha_{4.8}^{8.4}$ and at 98\% confidence level for 
$\alpha_{8.4}^{22}$. 

Although the evidence of steeper spectra for BAL
QSOs is at best marginal, with 91\% significance for the 
test between the combined BAL sample and the comparison 
sample at the frequency range 4.8-8.4 GHz, we can reject with a 
high confidence, above 97\%, that BAL QSOs have on average 
flatter radio spectra than non-BAL QSOs, both in the frequency 
ranges 4.8-8.4 GHz and 8.4-22 GHz. 
We interpret this result as statistical evidence
that the BAL QSOs in our sample, or in combination with MM08 sample, 
do not tend to have position angles 
closer to the radio axis than non-BAL QSOs, i.e. 
the orientation models for BAL QSOs in which 
they predominantly arise from polar winds 
(for instance Punsly 1999a, 1999b), 
are in contradiction with our results.
At a lower level of significance, the slightly steeper spectra 
of BALs compared to non-BAL in the range 4.8-8.4 GHz 
are consistent with the equatorial wind model of Elvis (2000).

Figure 5 shows histograms of the distributions of $\alpha_{4.8}^{8.4}$  
and $\alpha_{8.4}^{22}$ for the BAL and non-BAL QSO samples,
and statistics are presented in Table 10.
The spectral indices in the two frequency 
ranges show a mixture of flat ($\alpha \ge -0.5$) 
and steep ($\alpha < -0.5$) spectra for the BAL and non-BAL QSO samples 
(see also Table \ref{listalpha1}). 
The values found for BAL QSOs suggest that these QSOs are seen 
from a wide range of orientations with respect to the jet axis 
(both flat and steep sources are present). 
The same conclusion that BAL QSOs are not oriented
along a particular line of sight was obtained by Becker
et al. (2000) and MM08, also on the basis of the radio spectral
indices of BAL QSOs. 
The comparison in our work with a control sample similar in redshift, 
and radio and optical properties, except for the absence of broad 
absorption features, reveals a weak tendency for BAL QSOs to be on 
average steeper than non-BAL QSOs, and allow us to firmly conclude 
(above 97\% confidence) that
BAL QSOs do not have flatter spectra than non-BAL QSOs.
%

%
%
%
%


The two spectral-index distributions obtained for the FIRST-SDSS QSOs
in our work, and for the BAL QSO sample by MM08, can be compared to
those reported in the literature for other samples of radio QSOs.  A
useful comparison is with the B3-VLA QSO sample (Vigotti et al. 1997,
Vigotti et al. 1999), selected at 408 MHz, and complete down to
$S=100$ mJy, with multi-frequency data available from which we can
compute the spectral index in the range 4.8-10.6 GHz, close to the
range 4.8-8.4 GHz used in the present work.  The angular sizes at 1.4
GHz of the B3-VLA QSOs were measured from maps taken with the VLA in C
and D configurations (Vigotti et al. 1989) and for the most compact
sources from observations in A configuration (private communication).
Since most of the QSOs in our sample are unresolved, as well as
analysing the spectral-index distribution for the B3-VLA QSO sample as
a whole (largest angular size 131 arcsec), we also considered a
subsample with angular sizes below 2.3 arcsec, which is a
representative upper limit of the sizes of the unresolved sources in
our work (from 8.46-GHz VLA data).  

Statistics for the spectral
indices $\alpha_{4.8}^{10.6}$ of the B3-VLA QSOs are included in Table
\ref{spec_stat}, considering the whole sample and the sub-sample of
more compact sources, with sizes below 2.3 arcsec.  
A K-S test
shows that the spectral-index distributions of the compact B3-VLA and
BAL QSOs are similar at the 97\% confidence level, both for the sample
reported here and for the combined sample including MM08 BALs. The
comparison with non-BALs shows that B3-VLA QSOs are steeper, at a
99.1\% confidence level. There is no obvious physical reason for
B3-VLA QSOs to be more similar to BAL QSOs than to non-BAL QSOs, the
most plausible explanation being that the selection of B3-VLA QSOs at
a low frequency, 408 MHz, favours the inclusion of sources with steeper
spectra. The fact that the spectral indices of B3-VLA QSOs are more
similar to those of BAL QSOs than to those of non-BAL QSOs is a
consequence of the former being slightly steeper than the
latter. Furthermore, this result emphasizes the importance of using an
appropriate control sample.

The QSO sample from our work includes three sources classified as variable: 
the unresolved non-BALs 0029$-$09, 1005+48 and 1521+43 and the resolved 
BAL QSO 1603+30. 0029$-$09 and 1521+43 are the flatter sources in the total 
QSO sample, both with $\alpha_{4.8}^{8.4} > 0.5$. 
The resolved BAL QSO 1603+30 has a spectral index near the 
limit between flat and steep spectra, 
with $\alpha_{4.8}^{8.4} = -0.45$. 1005+48 shows modest variability 
(just at the adopted thresholds of significance and fractional variability) and its 
spectral index is $\alpha_{4.8}^{8.4} = -0.67$. 
This trend between variability and a flat radio spectra is consistent 
with the expectation that a flat spectrum source is more likely to present 
Doppler beaming, and therefore have any existing variability 
magnified due to this same effect. 

The rest-frame radio luminosities of the sources, $L_{\rm ~4.8 GHz}$ 
are listed in the penultimate column of Table \ref{listalpha1}. They were calculated using the total 
flux density at 4.86 GHz from the VLA or from Effelsberg if VLA data were 
not available from this work. The k-correction was obtained using the 
spectral index $\alpha_{4.8}^{8.4}$ listed in the same table. 
The radio luminosities range from $10^{26.1}$ to 
$10^{28.6}$, above the limit $L_{\rm rad}=10^{26}$ W Hz$^{-1}$ generally 
adopted for radio-loud QSOs (Miller et al. 1990).



\subsection{Polarisation}
   

\begin{table*}
 \centering
\caption{Fractional polarisation $m$ (in percentage), at several frequencies ($\nu$ in GHz), 
for the BAL QSO sample. 
When no polarisation measurement was possible at 1.4 GHz, the NVSS value was used 
(indicated with an asterisk). 
Last two columns show the rotation measure (observed and corrected values).} 
\label{pol1}
\renewcommand\tabcolsep{3pt}
\begin{tabular}{ccccccccccccc}
\hline
\multicolumn{1}{c}{Name}&  
\multicolumn{1}{c}{$m_{1.4}$} &  
\multicolumn{1}{c}{$m_{2.6}$} & 
\multicolumn{1}{c}{$m_{4.85}$} & 
\multicolumn{1}{c}{$m_{4.86}$} &
\multicolumn{1}{c}{$m_{8.3}$} & 
\multicolumn{1}{c}{$m_{8.46}$} & 
\multicolumn{1}{c}{$m_{10.5}$}& 
\multicolumn{1}{c}{$m_{22}$}  &  
\multicolumn{1}{c}{$m_{43}$}  &
\multicolumn{1}{c}{Observed RM}  	  &
\multicolumn{1}{c}{Rest-frame RM}  \\
& & & & & & & & & & \multicolumn{1}{c}{(rad m$^{-2}$)} & \multicolumn{1}{c}{(rad m$^{-2}$)}\\
\hline
0044+00   &1.2$\pm$0.9*   	  &-	      &-     &$<$3.8	 &-	     &$<$4.7	 &-&$<$74.3    &-&-	         &-	        \\
0756+37   &2.4$\pm$0.7\phantom{*} &1.8$\pm$0.5&-     &1.1$\pm$0.2&1.5$\pm$0.4&1.4$\pm$0.3&-&$<$10.1    &-&60.4$\pm$2.1   &643$\pm$26	\\
0816+48   &5.7$\pm$1.2\phantom{*} &-	      &-     &$<$4.3	 &-	     &$<$7.0	 &-&-          &-&-	         &-		\\
0842+06   &$<$20.5\phantom{*}	  &-	      &-     &$<$3.8	 &-	     &$<$6.2	 &-&-          &-&-	         &-		\\
0849+27   &$<$2.1*		  &-	      &-     &-      	 &-	     &-          &-&-          &-&-	         &-		\\
0905+02   &7.7$\pm$3.8\phantom{*} &-	      &-     &$<$4.7 	 &-	     &4.2$\pm$1.8&-&$<$61.7    &-&-	         &-		\\
0929+37   &6.1$\pm$0.9\phantom{*} &-	      &-     &$<$3.9	 &-	     &$<$4.1	 &-&$<$25.5    &-&-	         &- 		\\
1014+05   &$<$4.7\phantom{*}      &-	      &-     &2.9$\pm$1.1&-	     &$<$4.9	 &-&$<$49.5    &-&-	         &-		\\
1040+05   &$<$3.2*		  &-	      &-     &4.4$\pm$0.9&-	     &$<$24.6	 &-&-          &-&-	         &-		\\
1054+51   &4.2$\pm$1.1\phantom{*} &-	      &-     &-	         &-	     &$<$11.8	 &-&-	       &-&-	         & -		\\
1102+11   &1.9$\pm$0.9\phantom{*} &-	      &-     &$<$3.7 	 &-	     &$<$6.2 	 &-&$<$377.8   &-&-	         &-		\\
1103+11   &$<$1.2\phantom{*}	  &-	      &-     &$<$1.5 	 &-	     &$<$1.8	 &-&$<$13.1    &-&- 	         &-		\\	  
1129+44   &$<$7.5\phantom{*}	  &-	      &-     &$<$4.1	 &-	     &$<$4.7	 &-&-          &-&-	         &-		\\
1159+01   &6.5$\pm$0.7\phantom{*} &-	      &-     &1.9$\pm$0.3&-          &0.7$\pm$0.2&-&$<$4.9     &-&$-$79.2$\pm$1.8~~&$-$822$\pm$16~~~	\\
1159+06   &1.3$\pm$0.5\phantom{*} &-	      &-     &2.1$\pm$0.7&-	     &3.6$\pm$1.3&-&-          &-&151.1$\pm$2.1	 &1436$\pm$21~~	\\
1229+09   &3.1$\pm$1.4\phantom{*} &-	      &-     &-          &-	     &-       	 &-&-	       &-&-		 &-		\\
1237+47   &$<$3.7\phantom{*}	  &-	      &-     &$<$3.8 	 &-	     &5.9$\pm$2.6&-&-          &-&-		 &-		\\
1304+13   &$<$5.5\phantom{*}	  &-	      &-     &$<$5.9	 &-	     &-    	 &-&$<$77.7    &-&-		 &-		\\
1327+03   &$<$2.1\phantom{*}	  &$<$83.2    &-     &-   	 &-	     &$<$2.3	 &-&-          &-&-		 &-		\\
1335+02   &$<$2.5\phantom{*}	  &-	      &-     &$<$3.7 	 &-	     &$<$1.9	 &-&$<$4.2     &-&-		 &-		\\
1337$-$02 &7.8$\pm$2.4\phantom{*} &-	      &-     &-    	 &-	     &$<$8.0	 &-&-          &-&-		 &-		\\
1404+07   &$<$1.6\phantom{*}	  &-	      &-     &$<$5.3 	 &-	     &0.6$\pm$0.2&-&$<$3.9     &-&- 		 &-	        \\  	
1406+34   &2.9$\pm$0.9\phantom{*} &-	      &$<$4.3&1.0$\pm$0.2&1.7$\pm$0.2&3.5$\pm$0.2&-&2.3$\pm$1.1&-&284.0$\pm$4.5  &3520$\pm$57~~	\\
1603+30   &$<$2.4*		  &-	      &-     &$<$14.6	 &-	     &$<$4.1	 &-&$<$52.4    &-&-		 &-		\\
1624+37   &-\phantom{*}   	  &-	      &-     &-       	 &-	     &$<$17.9	 &-&-	       &-&-		 &$-$18350$\pm$570$^{\mathrm{1}}$~~~~\\
\hline
\end{tabular}
\begin{list}{}{}
\item[$^{\mathrm{1}}$] RM value taken from Benn et al. (2005). 
\end{list}

\end{table*}

\begin{table*}
 \centering
  \caption{Polarisation measurements 
for the sample of non-BAL QSOs (see the caption of Table \ref{pol1} for details). }
\label{pol2}
\renewcommand\tabcolsep{3pt}
\begin{tabular}{lcccccccccccc}
\hline
\multicolumn{1}{c}{Name}         &  
\multicolumn{1}{c}{$m_{1.4}$}    &  
\multicolumn{1}{c}{$m_{2.6}$}    & 
\multicolumn{1}{c}{$m_{4.85}$}   & 
\multicolumn{1}{c}{$m_{4.86}$}   &
\multicolumn{1}{c}{$m_{8.3}$}    & 
\multicolumn{1}{c}{$m_{8.46}$}   & 
\multicolumn{1}{c}{$m_{10.5}$}   & 
\multicolumn{1}{c}{$m_{22}$}     &  
\multicolumn{1}{c}{$m_{43}$}     &
\multicolumn{1}{c}{Observed RM}  	 &
\multicolumn{1}{c}{Rest-frame RM} \\

& & & & & & & & & & \multicolumn{1}{c}{[rad m$^{-2}$]} & \multicolumn{1}{c}{[rad m$^{-2}$]}\\
\hline
0014+01       	&7.8$\pm$2.4\phantom{*}&-            &-           & 9.5$\pm$3.1 &-	        &11.8$\pm$3.7~~ &-           &-           &-	&$-$7.8$\pm$0.6  &	8.7$\pm$6.1	\\
0029$-$09       &$<$6.9\phantom{*}       &-            &-           &$<$3.5 	  &-            & 1.6$\pm$0.7 &-           &$<$8.4      &$<$14.4	&-  &	-  	\\
0033$-$00       &3.8$\pm$0.9\phantom{*}	 &-            &-           &$<$6.8       &-            &$<$9.1       &-           &-           &-&-   &-   	\\
0103$-$11       &$<$2.3\phantom{*}       &-            &-           &$<$2.9	  &-            &$<$2.7	      &	-          &-           &     $<$47.6	&   	- 		&	- 		\\
0124+00       	&$<$1.2\phantom{*}       &11.1$\pm$2.1 &5.7 $\pm$1.1& 5.5$\pm$0.6 &-            & 4.5$\pm$1.1 &	-          &$<$38.1     &     -	&       43.7$\pm$7.9	& 	257$\pm$64~~	\\
0125$-$00       &1.2$\pm$0.1*            &-            &4.5 $\pm$0.9& 6.4$\pm$0.3 &-            & 6.9$\pm$0.3 &5.6$\pm$1.0 &4.8$\pm$1.8	&     $<$14.2	&	108.2$\pm$4.2~~  	&	1041$\pm$45~~~~	\\
0125$-$00A$^1$	&-	                 &-            &-           &-            &-            &11.0$\pm$0.8~~ &-           &-           &-   &-		&-\\
0125$-$00C$^1$	&-                       &-            &-           &-            &-            & 6.3$\pm$0.2 &-	   &-           &-&-&	-	\\
0152+01       	&$<$3.8*	         &-            &-           &24.6$\pm$5.8~~ &-            &$<$14.9      &-       	   &-           &     -	&   	-		&-		\\
0154$-$00       &$<$1.1\phantom{*}       &-            &-           & 1.2$\pm$0.3 &-            &-            &-	   &$<$16.6     &     -	&	-		&	-		\\
0158$-$00       &3.5$\pm$0.4\phantom{*}  &-            &-           & 1.9$\pm$0.5 &-	        &$<$2.4	      &-           &$<$18.4     &     -	& 	-&	-		\\
0750+36       	&$<$11.9\phantom{*}      &-            &-           & 1.4$\pm$0.7 &-	        &$<$2.6	      &-           &$<$26.5     &     -     	&-&	-		\\
1005+48         &9.1$\pm$1.9\phantom{*}  &11.9$\pm$0.8 &$<$95.1     &11.3$\pm$0.4~~ &11.4$\pm$0.9~~ &12.4$\pm$0.8~~ &12.1$\pm$3.0~~&15.4$\pm$3.4~~&$<$56.8&~~5.1$\pm$0.7&$-$26.9$\pm$8.0~~~~	\\
1322+50       	&2.1$\pm$0.3\phantom{*}  &-            &-           &$<$2.6       &-            &$<$3.5       &-           &$<$33.1	&   -	&-  &	-  \\
1333+47         &2.3$\pm$0.6\phantom{*}  &-            &-           &$<$5.2       &-            &-            &-           &$<$59.6	&   -     &-&-\\
1401+52         &7.0$\pm$1.5\phantom{*}  &-            &-           &$<$3.5       &-            & 4.8$\pm$1.6 &-           &$<$59.3	&    -   &-&-\\
1411+34       	&5.0$\pm$0.3*            & 5.6$\pm$1.1 &-           & 5.9$\pm$0.7 &-            & 4.8$\pm$0.6 &-           &$<$3.0      &     $<$1.0&   $-$238$\pm$16~~ &	$-$1937$\pm$130~~~~	\\
1411+43       	&2.2$\pm$0.6\phantom{*}  &-            &-           & 1.2$\pm$0.3 &-            & 2.2$\pm$0.5 &-           &$<$8.0      &     $<$29.8&   $-$174.4$\pm$7.4~~~~	&	$-$3077$\pm$130~~~~ \\
1502+55       	&4.0$\pm$0.9\phantom{*}  &-            &-           & 5.6$\pm$1.2 &-            & 6.3$\pm$1.9 &-           &$<$69.6     &     -&96.6$\pm$0.6	&1668$\pm$11~~~~ 	\\	  	
1512+35       	&$<$2.2*                 &-            &-           & 1.8$\pm$0.9 &-            & 2.4$\pm$0.9 &-           &$<$10.9	&     -	&-&	-		\\
1521+43       	&1.6$\pm$0.3\phantom{*}  &-            &-           & 0.7$\pm$0.2 &-            & 2.6$\pm$0.1 &-           &1.2$\pm$0.5	&   -	&$-$72.1$\pm$5.3~~   	&	$-$725$\pm$53~~~~	\\
1528+53       	&$<$2.4\phantom{*}       &-            &-           &-            &-            & 4.5$\pm$1.1 &-           &$<$27.4	&     -	&-   	&	-\\
1554+30         &1.9$\pm$1.1*  	         &-            &-           &-            &-            &$<$17.8      &-           &$<$87.6	&    -      	&	-&-		\\
1634+32       	&$<$7.3\phantom{*}       &-            &1.5$\pm$0.4 & 1.7$\pm$0.2 &-            & 2.8$\pm$0.3 &-&$<$18.4   &$<$38.7     &281$\pm$40  	&2950$\pm$450~~	\\
1636+35       	&-\phantom{*}	         &-            &-	    & 2.8$\pm$0.5 & 5.9$\pm$1.0 &-            &-           &$<$18.6     &     -&	-   	&	-		\\
1641+33       	&2.1$\pm$1.0\phantom{*}	 &-            &-           & 2.1$\pm$0.5 &-            & 3.1$\pm$0.7 &-           &$<$28.0	&  -&23.2$\pm$0.1    &107.1$\pm$1.4~~~~  	\\
1728+56       	&2.3$\pm$0.6\phantom{*}	 &-            &-           & 3.2$\pm$0.6 &-            &$<$3.7       &-           &$<$50.8	&    -	&	-   	&	- \\	  	
2109$-$07       &3.6$\pm$1.4\phantom{*}	 &-            &-           & 4.2$\pm$1.3 &-            &$<$7.2	      &-           &-           &     -	&	-   	&	- \\
2129+00         &$<$4.3\phantom{*}       &-            &-           &$<$3.0       &-            &$<$5.4       &-           &-           & - & -	&-\\
2143+00       	&-\phantom{*}            &-            &-           &$<$2.3 	  &-            &$<$2.2	      &-           &$<$15.7	&    $<$20.2	&-   	&-\\
2238+00       	&$<$7.6\phantom{*}       &-            &-           &-            &-            &-            &-           &-     	&    -&-	&	-\\
2244+00       	&$<$8.0\phantom{*}       &-            &-           &$<$4.2       &-            &$<$6.2	      &-           &$<$52.6	&     -&   -   		&-\\
2248$-$09       &3.3$\pm$1.4*            &-            &-           &$<$7.7       &-            &$<$14.4      &-           &-           &   -      	&   -	&-\\
2331+01         &$<$3.7*                 &-            &-           &$<$4.5       &-            &$<$5.5       &-           &$<$37.6	&    -	&	-&-	\\
2346+00       	&6.8$\pm$1.8\phantom{*}	 &-            &-           &$<$5.1       &-            & 1.8$\pm$0.8 &-	   &$<$12.4     &     $<$39.7	&-   &	-\\
2353$-$00       &$<$3.3\phantom{*}       &-            &-           &$<$4.3       &-            &$<$6.6	      &-           &-           &    -	&   -	&-	\\
\hline
\end{tabular}
\begin{list}{}{}
\item[$^{\mathrm{1}}$] Measurements for a specific component (see Fig. 2). 
\end{list}

\end{table*}

For all sources we have obtained $S_Q$ and $S_U$  in order to calculate the fractional polarisation $m$ 
and the polarisation angle $\chi$. Most of the measurements were obtained from the VLA observations. 
 In only a few cases did the Effelsberg observations have high-enough
signal-to-noise
to detect polarisation 
fractions below 10\%.  Only 3-$\sigma$ results were considered, except for the cases for which a detection 
above 2-$\sigma$ resulted in a consistent $m$ with respect to the other frequencies, increasing the 
measurement reliability. Values of the fractional polarisation are presented in Tables \ref{pol1} 
and \ref{pol2} for BAL and non-BAL QSOs respectively. We have included the NVSS values (NRAO VLA Sky Survey, 
Condon et al. 1998) for the polarisation fraction at 1.4 GHz when no measurements could be obtained 
from our data. The polarisation measurements as well as the more significant upper limits were 
mostly obtained for frequencies in the range from 1.4 to 8.5 GHz.

\begin{figure}[tbp]
\centering
\includegraphics[width=100mm]{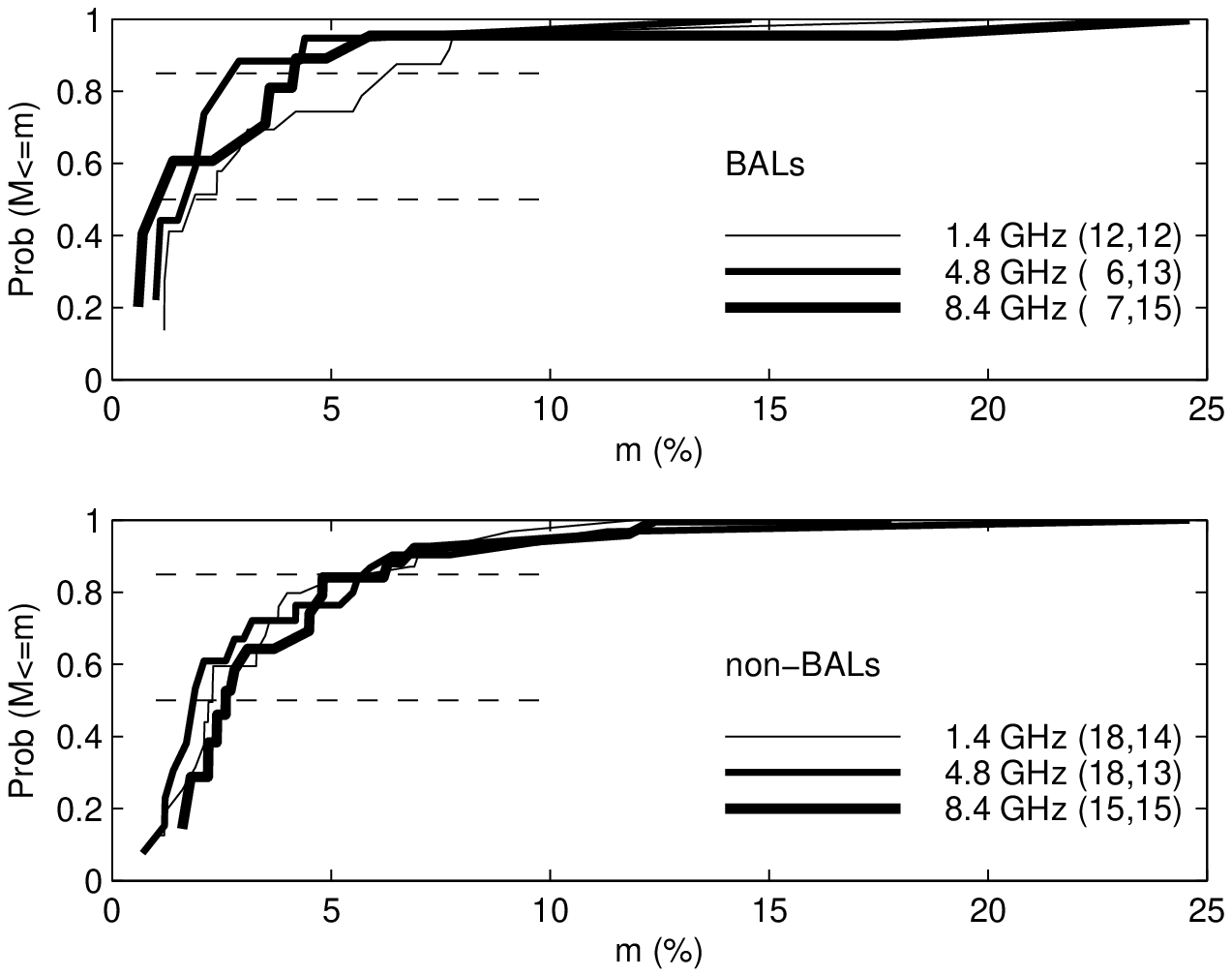}
\caption{
Cumulative distribution of the fractional polarisation for the 
two samples, at each of three frequencies. 
Each pair of numbers in parenthesis indicates the number of detections 
and the number 
of upper limits at that frequency. 
The dashed lines show the 50\% percentile (i.e. the median) and the 85\% percentile.}\label{KM}
\end{figure}

We have obtained the cumulative distribution function $F(m) = {\rm Prob} (M \le m)$  for the fractional 
polarisation at 1.4, 4.8 and 8.4 GHz for the two samples, using the Kaplan-Meier estimator, 
that allows inclusion of information from upper-limits. The method is described in detail in 
Feigelson \& Nelson (1985). The results are shown in Fig. \ref{KM}. The pair of numbers 
in parenthesis indicates the number of detections and the number of upper limits for each 
frequency.

The distributions for the non-BAL QSOs are very similar for the three frequencies, 
yielding a median value $m$ in the range 1.8-2.5\%. 
The 85\% percentile corresponds to $m \le 5.8-6.3\%$. 
Table \ref{pol2} shows five sources exceeding $m = 10\%$ at some frequency.
These sources are 0014+01, 0124+00, 0152+01, 1005+48 and the lobe component of 0125$-$00. 
In particular 1005+48 shows fractional polarisation above 10\% at a wide range of frequencies, 
from 2.6 to 22 GHz. 
The distributions for the BAL QSOs have a higher uncertainty, due to the fewer data 
and the larger fraction of upper limits, especially at  4.8 and 8.4 GHz.  
From the 1.4 GHz data we find a median $m = 1.8\%$ and a limit  $m \le 6.2\%$ for the 85\% percentile. 
None of the measurements in Table \ref{pol1} is above $m=10\%$.

Regarding the three BAL QSOs in common with MM08, we note that for 1603+30 these authors obtained 
$m=1\%$ at 8.4 GHz and upper limits for the remaining frequencies in their study. 
For this source we only obtained upper limits, and the one at 8.4 GHz is consistent 
with the measurement at MM08.  For 1159+01 our measurements at 4.8 and 8.4 GHz agree 
with the results by MM08, although at 1.4 GHz we obtained $m=6 \%$, less than half 
the value reported by MM08, of $m=15\%$ (taken from NVSS). For the remaining source, 
1624+37, our data only provide a high upper limit for $m$, but the source shows a high 
fractional polarisation from the data reported in Benn et al. (2005, their Table 2), 
with $m=6\%$ at 4.8 GHz and $m=11\%$ at 10 and 22 GHz.

The fractional polarisations of BAL and non-BAL QSOs appear to be similar, 
with median values around 1-3\%, 85\% of the sources having $m < 6\%$, 
and around ten per cent of the sources showing fractional polarisation 
above 10\% at some frequency (4/34 for the non-BAL QSOs and 2/25 for the BAL QSOs, considering the information from the literature).

An extensive survey of the fractional polarisation of QSOs is
presented in Pollack et al. (2003), based on a sample with 
$S_{\rm 4.85~GHz} \ge 350$ mJy and radio spectral index $\alpha_{1.4}^{4.8} \ge
-0.5$. The authors compute $m$ at 4.85 GHz separately for the core and
jet components, and find a higher polarisation for the jet
components. In particular, the 85\% percentile corresponds to $m\le3\%$
for the core components and to $m\le14\%$ for the jet components, and
the proportion of sources exceeding a fractional polarisation of 10\% 
is 1/91 for the cores and 17/43 for the jet components. The fractional 
polarisations we found for the BAL and non-BAL QSOs in our sample occupy
an intermediate range between the results found by Pollack et
al. (2003) for core and jet components.

Sadler et al. (2006) obtained the polarisation fraction or upper
limits for 41 QSOs in their sample of radio sources selected at 20
GHz, with $S_{\rm 20~GHz} \ge 100$ mJy.  As mentioned in Section 4.2,
this QSO sub-sample is dominated by flat-spectrum sources (69\% with
$\alpha_5^8 \ge -0.5$, compared to 42\% for the SDSS-FIRST QSOs in our
sample). We obtained the cumulative distribution function of $m$ for
this sample, using the Kaplan-Meier estimator, and found $m = 2.4 \%$
and $m \le 4.2\%$ for the median and the 85\% percentile
respectively. None of the QSOs in the Sadler et al. (2006) sample exceed
polarisation levels of 10\%. The results of Sadler et al. (2006) show good
agreement with those of Pollack et al. (2003) for the core components.

\subsection{Rotation measures}

With at least three measurements of the polarisation angle $\chi$ at 
different frequencies it is possible to estimate the rotation 
measure (RM) of a source, via a linear fit of the 
polarisation angle versus the square of the observed wavelength 
$\lambda$ ($\chi = \chi_0 + {\rm RM } \lambda^2$, $\chi_0$ being 
the intrinsic polarisation angle). The rotation measure, 
which is the slope of the fit, is proportional to the magnetic field 
component along the line of sight, to the electron density, 
and to the path length,

\begin{equation}
{\rm RM} \propto \int_0^L n_e ~ B_{\parallel} ~ dl  
\end{equation}

Plots of the linear fits for the four BAL QSOs and ten non-BAL QSOs with at least three 
measurements are shown in Figure 7. 
The observed RM values are listed in Tables \ref{pol1} and \ref{pol2}, 
along with the RM values corrected from the Galaxy contribution and converted to the rest-frame, 
multiplying by the factor $(1+z)^2$. 
Since the sources are located well above the Galactic Plane ($b > 28^{\circ}$) 
the applied Galactic correction was small, in the range from $-9$ to 17 rad m$^{-2}$ 
(Taylor et al. 2009). 

There is good agreement between the observed rotation measure for 1159+01 in this work, 
of 79.2$\pm$1.8 and the result from MM08, of 72.1$\pm$1.4. 
The RM listed in Table \ref{pol1} for 1624+37 was taken from Benn et al. (2005). 
With a rest-frame RM of 18350 $\pm$ 570 rad m$^{-2}$, this was and still is the second-highest 
RM known, after that of quasar OQ172 (Kato et al. 1987; O'Dea 1998), with 
RM = 22400 rad m$^{-2}$.

For the non-BAL QSO sample we have data available for ten sources, allowing 
some
statistical analyis.  We found rest-frame $|{\rm RM}|$ values in the range from 8.7 to 3077 rad m$^{-2}$, 
with a median value of 883 rad m$^{-2}$, and average and standard deviation of 1180 rad m$^{-2}$ and 1117 rad m$^{-2}$. 
Three of the BAL QSOs have rest-frame $|{\rm RM}|$ values within the range found for the non-BAL QSOs. 
One of the remaining BAL QSOs with available rotation measure is 1406+34, with RM = 3520 $\pm$ 57 rad m$^{-2}$, 
which is only 2$\sigma$ from the average value for the non-BAL QSOs. The other source is 1624+37,  
with the extremely high value RM = $-18350 \pm 570$. 
The small number of data for BAL QSOs does not allow to compare the rotation measures of the two samples.

Zavala \& Taylor (2004) reported rest-frame rotation measures, $|{\rm RM}|$, 
for the flat core components of 26 QSOs in the range from 200 to 10000 rad m$^{-2}$, 
with a median value of 1862 rad m$^{-2}$ and average 2515 rad m$^{-2}$. 
The same authors found lower values for the rotation measures of the steep jet components of these QSOs, 
for which they obtained a median value of 458 rad m$^{-2}$ and average 600 rad m$^{-2}$. 
The statistics for the non-BAL QSOs in our sample places them in the intermediate 
range of rotation measures between those of the flat and steep components in Zavala \& Taylor (2004) sample.


\section{Conclusions}

We constructed a sample of 59 radio-loud QSOs with $S_{1.4} > 30$ mJy,
selected by cross-correlating the FIRST radio survey and the 
4th edition of the SDSS Quasar Catalogue, at redshifts such that the 
wavelength range from \ion{Si}{iv} 1400 {\AA} to \ion{C}{iv} 1550 {\AA} is covered by the 
SDSS spectra. The sample comprises 25 sources having definite broad
absorption in \ion{C}{iv} with velocity width of at least 1000 km s$^{-1}$,
referred to as the ``BAL QSO sample'', 
and 34 sources lacking such absorption, which
form the ``non-BAL QSO comparison sample''.  
The sources were observed at 
frequencies ranging from 1.4 to 43 GHz, using the 100-m Effelsberg
telescope and at the VLA, and these have allowed us to compare several
radio
properties of the two samples, 
including morphology, 
flux-density variability, 
spectral shapes,
spectral-index distributions, and polarisation properties:

\begin{figure*}[tbp]
\centering
  \includegraphics[width=150mm]{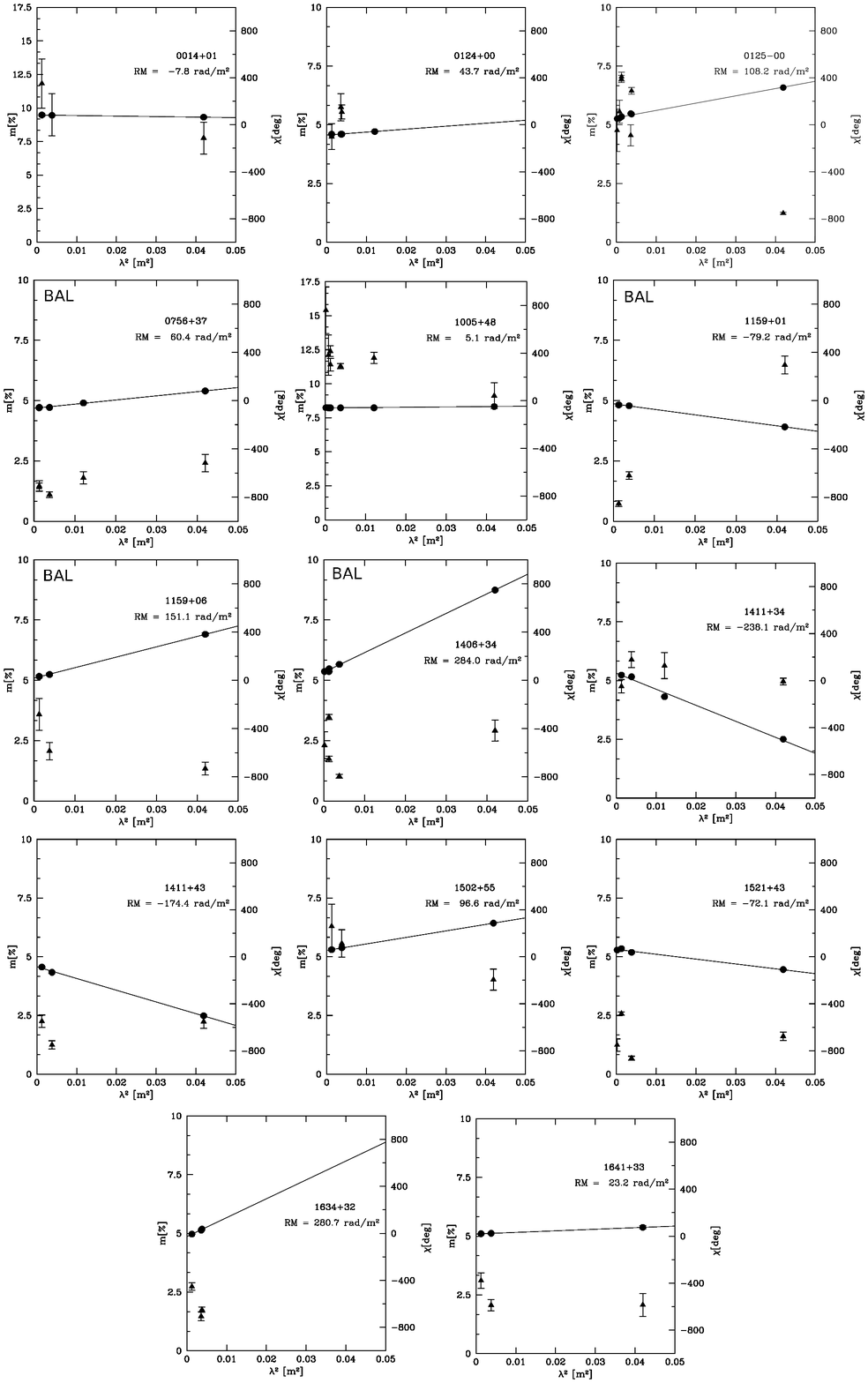}
\caption{
Linear fits of the polarisation angles $\chi$ versus the square of the observed wavelengths 
for 4 BAL QSOs and 10 non-BAL QSOs, yielding the rotation measures. 
The errors in the position angles are lower than the dot size. 
Triangles correspond  to the polarisation percentages ($m$).}\label{RM_BALs}
\end{figure*}

\subsection{Linear sizes}
Only eight of the 59 sources are extended at the arcsec level,
four of them being BAL QSOs and four of them non-BAL QSOs. 
The fractions of resolved sources are similar for
BALs (16\%) and  non-BALs (12\%),
and the distributions of linear sizes are also similar, ranging 20 to 200-400 kpc. 
The morphologies 
are also similar, including elongated sources (1), core-lobe 
(4, possibly 5) and core double-lobe (2, possibly 3). 
90-95\% of the unresolved sources have an estimated size at 8.46 GHz 
below 20 kpc, obtained 
from VLA observations at this frequency (2.3 arcsec resolution),
and adopting the average redshift $z=2.4$ for the two samples. 
Two of the unresolved BAL QSOs, 1159+01 and 1624+37, were resolved from VLBA 
observations at Montenegro-Montes et al. (2008b, 2012 in preparation), both 
showing a core-jet morphology with sizes of 0.85 kpc and 60 pc respectively.\\

\subsection{Flux-density variability}
The flux-density variability of the sources at 4.8 and 8.4 GHz was computed 
from observations at VLA and Effelsberg, at the same frequency, 
in two epochs separated typically by 1.6 years. 
In addition, for the 
3 BAL QSOs in common with MM08, the flux densities from our work at 
various frequencies in the interval 2.6-22 GHz (VLA or Effelsberg) were compared 
to similar data (frequency and telescope) from MM08, with a time difference of 3-4 years. 
Excluding variations that could be attributed to resolution
effects (higher flux at the lower angular resolution for any
source, whether resolved or not), we found three likely variables 
from our data, all non BALs (0029-09, 1005+48, 1521+43). 
Using flux densities from the literature, we concluded that the BAL QSO 
in our sample 1603+30 was also a candidate variable. 
Our data suggest a lower rate of variable sources among BALs, 
compared to non-BALs. However, this needs to be confirmed by a 
monitoring of the sample at various epochs using the same configuration 
for the radio observations.

The proportion of variable sources exceeding a fractional
variability of 20\% in our total sample, 4/59, is consistent
with the fraction for the core-dominated QSO
sample of Barvainis et al. (2005), selected at 8 GHz and yielding 
a fraction of 5/50 variables at this frequency. The
proportion is also consistent with the fraction 2/32 
from the QSO sample studied at 20 GHz by Sadler et al. (2006).
\\ 

\subsection{Radio spectral shape}
We found 9 BAL QSOs and 8 non-BAL QSOs with GPS-like radio 
spectra, having peak frequencies in the range from 0.5 to 
7 GHz in the observer frame.  
Their linear sizes (or limits thereon, typically 20 kpc at 8 GHz) 
are consistent with a maximum allowed size of 1 kpc, for classification 
as a GPS source.
Higher-resolution observations are needed to confirm or reject 
the GPS classification of these sources. In particular, 
this classification is confirmed for 1624+37, 
with a size of 60 pc at 5 GHz and 75 pc at 8 GHz 
(Montenegro-Montes et al. 2008b, 2012 in preparation).

The fractions of candidate GPS sources are
$36 \pm 12\%$ (9/25) for BAL QSOs and $23 \pm
8\%$ (8/34) for non-BAL QSOs, i.e. no significant difference.  
Given the
widespread interpretation that GPS are young sources, our result
suggests that BAL QSOs are not a younger population than non-BAL QSOs.

Low-frequency upturns in some of the radio spectra indicate additional 
low frequency components, in about 
$12 \pm 7$\% of the BAL QSOs (3/25) and $15 \pm 7$\% of the non-BAL QSOs
(5/34). The two values are similar within the errors, and since the  
low frequency excess emission likely corresponds to old components, 
this result again favours similar 
distributions of ages for BAL and non-BAL QSOs.\\

\subsection{Radio spectral indices}
We found a mix of flat ($\alpha \ge -0.5$) and steep ($\alpha < -0.5$) spectra 
for both the BAL and the non-BAL QSO samples, suggesting that both classes 
are seen from a wide range of orientations with respect to the jet axis. 
A similar conclusion was reached by Becker et al. (2000) and MM08, 
on the basis of the radio spectral-index distribution of BAL QSOs. 
Kolmogorov-Smirnov tests comparing the spectral index 
distribution $\alpha_{4.8}^{8.4}$ of the two samples 
provide weak evidence (at the 91\% confidence level) 
that the spectra in the combined BAL QSO sample from our work and MM08 
are steeper than those in the non-BAL QSO sample, 
and significant evidence ($\ge$ 97\% confidence) 
that both our BAL sample and the combined BAL QSO sample 
are not flatter than the non-BAL QSO sample.
The latter result indicates that 
radio-loud BAL QSOs do not tend to have position angles closer to the 
radio axis than non-BAL radio-loud QSOs, 
i.e. a model in which the BAL absorption arises 
predominantly from polar winds (for instance Punsly 1999a, 1999b), 
is not consistent with our results.\\


\subsection{Radio polarisation properties}
The fractional polarisations $m$ of the BAL and non-BAL QSOs are
similar, median 1-3\%, with 
$\sim$ 85\% of the sources having $m < 6$\%, and $\sim$ 10\% 
having $m >$ 10\% at some
frequency. 
These values are intermediate between the
values found by Pollack et al. (2003) for the core and jet 
components of the QSOs in their sample, with higher 
values for the latter.

The Rotation Measure has been determined for 5 BAL QSOs
and 10 non-BAL QSOs. 
The Rotation Measures for the non-BALs range 9 to 3100 
rad m$^{-2}$, with mean and standard deviation 
1180 $\pm$ 1120 rad m$^{-2}$, intermediate between flat- and steep-spectrum 
(higher RM) components in the sample of QSOs of Zavala \& Taylor (2004). 
The limited statistics provide no evidence for a significant difference
between the RM of BAL and non-BAL QSOs.
The only BAL QSO exceeding by more than 2$\sigma$ the 
mean value found for non-BAL QSOs 
is 1624+37, with an unusually high 
$\rm{RM} = 18350 \pm 570$ rad m$^{-2}$.

\subsection{Summary}
We have compared the distributions of linear size, flux-density variation,
spectral shape, spectral index and polarisation properties for 
samples of BAL and non-BAL radio QSOs.

We find these distributions to be statistically indistinguishable,
except for weak evidence that the spectra of BAL QSOs are steeper than
those of non-BALs.  The latter difference 
mildly favours edge-on orientations for BAL QSOs, but the spectral indices are
still consistent with a broad range of orientations.

At a high level of significance, we can exclude the possibility that
the spectra of BAL QSOs are 
flatter than those of non-BAL QSOs, ruling out a preferred polar
orientation for the former.

The similarity of the fractions of GHz-peaked sources in the two samples 
suggests that BAL QSOs are not generally younger than non-BAL QSOs.


\begin{acknowledgements}
We are grateful to F. Mantovani for helping us during the observations
at the 100-m Effelsberg telescope.  We would like to thank also
A. Mignano, from the ALMA Regional Centre (Italian node), for helping
us in image analysis with the first release of the CASA astronomical
software.\\
Part of this work was supported by a grant of the Italian Programme for
Research of Relevant National Interest (PRIN No. 18/2007, PI: K.-H. Mack)
%
The authors acknowledge financial support from the Spanish Ministerio de Ciencia 
e Innovaci\'on under project AYA2008-06311-C02-02.\\
This work has benefited from research funding from the European Union's 
sixth Framework Programme under RadioNet grant agreement no. 227290.\\ 
This work has been partially based on observations with the 100-m 
telescope of the MPIfR (Max-Planck-Institut f\"ur Radioastronomie) at Effelsberg.\\ 
The National Radio Astronomy Observatory is a facility of the National Science Foundation operated under cooperative 
agreement by Associated Universities, Inc.\\ 
This research has made use of the NASA/IPAC Infrared Science Archive and NASA/IPAC Extragalactic Database (NED) 
which are both operated by the Jet Propulsion Laboratory, 
California Institute of Technology, under contract with the National Aeronautics and Space Administration.\\ 
Use has been made of the Sloan Digital Sky Survey (SDSS) Archive. The SDSS is managed by the Astrophysical Research Consortium (ARC) for the participating institutions: The University of Chicago, Fermilab, the Institute for Advanced Study, the Japan Participation
Group, The John Hopkins University, Los Alamos National Laboratory, the Max-Planck-Institute for Astronomy (MPIA), the 
Max-Planck-Institute for Astrophysics (MPA), New Mexico State University, University of Pittsburgh, Princeton University,
the United States Naval Observatory, and the University of Washington.\\ 
\end{acknowledgements}




\begin{thebibliography}{99}
\bibitem[\protect\citeauthoryear{Adelman-McCarthy et al.}{2007}]{Adelman} Adelman-McCarthy, J.K., Agüeros, M.A., Allam, S.S. et al. 2007, AJ Supplement Series, 172, 2, 634
\bibitem[\protect\citeauthoryear{Baars et al.}{1977}]{Baars} Baars, J.W.M., Genzel,
R., Pauliny-Toth, I.I.K. et al. 1977, A\&A, 61, 99
\bibitem[\protect\citeauthoryear{Barvainis et al.}{2005}]{Barvainis} 
Barvainis, R., Leh\'ar, J., Birkinshaw, M. et al. 2005, ApJ, 618, 108
%
\bibitem[\protect\citeauthoryear{Becker et al.}{1995}]{Becker}Becker, R. H., White, R. L. \& Helfand, D. J. 1995, ApJ,
450, 559

\bibitem[\protect\citeauthoryear{Becker et al.}{2000}]{Becker1} Becker, R.H., White, R.L., Gregg, M.D. et al. 2000, ApJ, 538, 72
%
\bibitem[\protect\citeauthoryear{Becker et al.}{2001}]{Becker2} Becker, R.H., White, R.L., Gregg, M.D. et al.
2001, ApJS, 135, 227
%
\bibitem[\protect\citeauthoryear{Benn et al.}{2005}]{Benn} Benn, C.R., Carballo, R., Holt, J. et al. 2005, MNRAS, 360, 1455
%
\bibitem[\protect\citeauthoryear{Briggs et al.}{1984}]{Briggs} Briggs, F. H., Turnsheck, D. A. \& 
Wolfe M. 1984, ApJ, 287, 549 
%
\bibitem[\protect\citeauthoryear{Cohen et al.}{2007}]{Cohen} Cohen, A.S., Lane, W.M., Cotton, W.D. et al. 2007, AJ, 134, 1245
\bibitem[\protect\citeauthoryear{Condon et al.}{1998}]{Condon} Condon, J. J., Cotton, W. D., Greisen, E.W. et al. 1998, AJ, 115, 1693
\bibitem[\protect\citeauthoryear{Dallacasa et al.}{2000}]{Dallacasa} Dallacasa, D., Stanghellini, C., Centoza, M. et al. 2000, A\&A, 363, 887
\bibitem[\protect\citeauthoryear{de Bruyn et al.}{2000}]{Bruyn} de Bruyn, G., Miley, G., Rengelink, R. et al. 2000, VizieR On-line Data Catalog VIII/62
\bibitem[\protect\citeauthoryear{Douglas et al.}{1996}]{Douglas} Douglas, J.N., Bash, F.N. \& Bozyan, F.A. 1996, AJ, 111, 5, 1945
\bibitem[\protect\citeauthoryear{Elvis}{2000}]{Elvis} Elvis, M. 2000, ApJ, 545, 63
\bibitem[\protect\citeauthoryear{Fanti et al.}{1990}]{Fanti} Fanti, R., Fanti, C., Schilizzi, R. T. et al. 1990, A\&A, 231, 333
\bibitem[\protect\citeauthoryear{Feigelson \& Nelson}{1985}]{Feigelson} Feigelson, E. D. \& Nelson, P. I. 1985, ApJ, 293, 192
\bibitem[\protect\citeauthoryear{Ficarra et al.}{1984}]{Ficarra} Ficarra, A., Grueff, G. \& Tomasetti, G. 1984, A\&AS, 59, 255
\bibitem[\protect\citeauthoryear{Ghosh \& Punsly}{2007}]{Ghosh} Ghosh, K.K. \& Punsly, B. 2007, ApJ, 661, 139
\bibitem[\protect\citeauthoryear{Gupta et al.}{2006}]{Gupta} Gupta, N., Salter, C. J., Saikia, D. J., Ghosh, T., Jeyakumar, S. 2006, MNRAS, 373, 972
\bibitem[\protect\citeauthoryear{Hewett \& Foltz}{2003}]{Hewett} Hewett, P.C. \& Foltz, C.B. 2003, AJ, 125, 1784
\bibitem[\protect\citeauthoryear{Hales et al.}{1988}]{Hales} Hales, S.E.G., Baldwin, J.E. \& Warner, P.J. 1988, MNRAS, 234, 919
\bibitem[\protect\citeauthoryear{Hall et al.}{2002}]{Hall} Hall, P. B., Anderson, S. F., Strauss, M. A. et al. 2002, ApJS, 141, 267
\bibitem[\protect\citeauthoryear{Kato et al.}{1987}]{Kato} Kato, T., Tabara, H., Inoue, M. et al. 1987, Nat, 329, 223
\bibitem[\protect\citeauthoryear{Klein et al.}{2003}]{Klein} Klein, U., Mack, K.-H., Gregorini, L. et al. 2003, A\&A, 406, 579
\bibitem[\protect\citeauthoryear{Kovalev}{1995}]{Kovalev} Kovalev, Y. Y. 1996, in Extragalactic radio sources, ed. R. D. Ekers, C. Fanti, \& L. Padrielli., IAU Symp. 175,  Kluwer, Dordrecht, p. 95.
\bibitem[\protect\citeauthoryear{L\'ipari \& Terlevich}{2006}]{Lipari} L\'ipari, S.L. \& Terlevich, R.J. 2006, MNRAS, 368, 1011
\bibitem[\protect\citeauthoryear{Miller}{1990}]{Miller} Miller, L., Peacock, J.A. \& Mead, A.R.G. 1990, MNRAS, 244, 207
\bibitem[\protect\citeauthoryear{Montenegro-Montes et al.}{2008a}]{Montenegro} Montenegro-Montes, F.M., Mack, K.-H., Vigotti, M. et al. 
2008a, MNRAS, 388, 1853 (MM08)
\bibitem[\protect\citeauthoryear{Montenegro-Montes et al.}{2008b}]{Montenegro2} Montenegro-Montes, F.M., Mack, K.-H., 
Benn, C. et al. 2008b, PoS (IX EVN Symposium) 019 
\bibitem[\protect\citeauthoryear{Montenegro-Montes et al.}{2012}]{Montenegro3} Montenegro-Montes, F.M., Mack, K.-H., Dallacasa, D. et al. 2012 (in prep.)
\bibitem[\protect\citeauthoryear{Morganti}{2008}]{Morganti}Morganti, R. 2008, in Extragalactic Jets: Theory and Observation from Radio to Gamma Ray, ed. Rector, T. A., De Young, D. S., Girdwood, Alaska, Astronomical Society of the Pacific Conference Series, Vol. 386, p. 210
\bibitem[\protect\citeauthoryear{O'Dea \& Baum}{1997}]{ODea2} O'Dea, C.P \& Baum, S.A. 1997, AJ, 113, 148.
\bibitem[\protect\citeauthoryear{O'Dea}{1998}]{ODea} O'Dea, C.P. 1998, PASP, 110, 493
\bibitem[\protect\citeauthoryear{Orr \& Browne}{1982}]{Orr} Orr, M.J.L. \& Browne, I.W.A. 1982, MNRAS, 200, 1067
%
\bibitem[\protect\citeauthoryear{Punsly}{1999a}]{Punsly} Punsly, B. 1999a, ApJ, 527, 609
%
\bibitem[\protect\citeauthoryear{Punsly}{1999b}]{Punsly} Punsly, B. 1999b, ApJ, 527, 624
%
%
\bibitem[\protect\citeauthoryear{Pollack}{2003}]{Pollack} 
Pollack, L.K., Taylor, G. B. \& Zavala, R. T., 2003, ApJ, 589, 733
%
\bibitem[\protect\citeauthoryear{Sadler}{2006}]{Sadler} Sadler, E. M., Ricci, R., Ekers, R. D. et al. 2006, MNRAS, 371, 898
\bibitem[\protect\citeauthoryear{Schneider et al.}{2007}]{Schneider07} Schneider, D.P., Hall, P. B., Richards, G. T. et al 
2007, AJ, 134, 102
\bibitem[\protect\citeauthoryear{Stocke et al.}{1992}]{Stocke} Stocke, J.T., Morris, S.L., Weymann, R.J. et al. 1992, ApJ, 396, 487
\bibitem[\protect\citeauthoryear{Taylor et al.}{2009}]{Taylor} Taylor, A.R., Stil, J.M. \& Sunstrum, C., 2009 ApJ 702, 1230
\bibitem[\protect\citeauthoryear{Tornianien et al.}{2005}]{Tornianien} Torniainen, H., Tornikoski, M., Terasranta, H. et al. 2005, A\&A, 435, 839
\bibitem[\protect\citeauthoryear{Trump et al.}{2006}]{Trump} Trump, J. R., Hall, P. B., Reichard, T. A. et al. 2006, ApJS, 165, 1
\bibitem[\protect\citeauthoryear{van Breugel et al.}{1984}]{Breugel} van Breugel, W., Miley, G. \& Heckman, T. 1984, AJ, 89, 5
\bibitem[\protect\citeauthoryear{Vigotti et al.}{1989}]{Vigotti0} Vigotti, M., Grueff, G., Perley, R. et al. 1989, AJ, 98, 419
\bibitem[\protect\citeauthoryear{Vigotti et al.}{1997}]{Vigotti1} Vigotti, M., Vettolani, G., Merighi, R. et al. 1997 A\&AS 123, 219
\bibitem[\protect\citeauthoryear{Vigotti et al.}{1999}]{Vigotti2} Vigotti, M., Gregorini, L., Klein, U. et al. 1999, VizieR On-line Data Catalog: J/A+AS/139/359
\bibitem[\protect\citeauthoryear{Weymann et al.}{1991}]{Weymann} Weymann, R.J., Morris, S.L., Foltz, C.B. et al. 1991, ApJ, 373, 23
\bibitem[\protect\citeauthoryear{White et al.}{1997}]{White} White, R. L., Becker, R. H., Gregg, M. D. et al. 1997, American Astronomical Society, 191st AAS Meeting, \#103.05; Bulletin of the American Astronomical Society, 29, 1373
%
\bibitem[\protect\citeauthoryear{Zavala \& Taylor}{2004}]{Zavala} Zavala, R. T. \& Taylor, G. B., 2004, ApJ, 612, 749
%
\bibitem[\protect\citeauthoryear{Zhou et al.}{2006}]{Zhou} Zhou, H., Wang, T., Wang, H. et al. 2006, ApJ, 639, 716


\end{thebibliography}
\end{document}